\documentclass[11pt,nopacs,preprintnumbers,superscriptaddress,amsmath,amssymb,aps,nofootinbib]{revtex4}
\usepackage{graphicx}% Include figure files
\usepackage{dcolumn}% Align table columns on decimal point
\usepackage{bm}% bold math
\usepackage{amssymb}
\usepackage{amsmath}
\usepackage{epsfig}
\usepackage{color}
\usepackage{slashed}
\usepackage{hhline}
\usepackage{ulem}
\usepackage{bbm}
\usepackage{hyperref}
\usepackage{tikz-feynman}
\tikzfeynmanset{compat=1.1.0}
\def\be{\begin{equation}}
\def\ee{\end{equation}}
\newcommand{\bea}{\begin{eqnarray}}
\newcommand{\eea}{\end{eqnarray}}
\newcommand{\nn}{\nonumber}

\allowdisplaybreaks

\newcommand{\modYS}[1]{\textcolor{magenta}{#1}}

\begin{document}

%{\begin{flushright}{APCTP Pre2023 - 0XX}\end{flushright}}

%%%%%%%%%
\title{A Minimal Realization of Radiative Dirac Neutrino Masses via a Non-Invertible Fusion Rule}
%\title{Radiative Dirac Neutrino Masses from a Minimal Non-Invertible Fusion Framework}
%\preprint{xxx}

\author{Takaaki Nomura}
\email{nomura@scu.edu.cn}
\affiliation{College of Physics, Sichuan University, Chengdu 610065, China}

\author{Hiroshi Okada}
\email{hiroshi3okada@htu.edu.cn}
\affiliation{Department of Physics, Henan Normal University, Xinxiang 453007, China}

\author{Yoshihiro Shigekami}
\email{shigekami@htu.edu.cn}
\affiliation{Department of Physics, Henan Normal University, Xinxiang 453007, China}

\date{\today}

\begin{abstract}
{
We propose a minimal one-loop radiative framework for Dirac neutrino mass matrix. 
As a consequence, the Yukawa hierarchies among the SM fermions are alleviated, and radiative type-I seesaw framework is realized. 
To regulate divergent loop contributions, we introduce an effective cutoff scale $\Lambda \sim 100 \, {\rm TeV}$. 
By introducing a scalar leptoquark and imposing appropriate assignments of ising fusion rule to the particle content, we successfully realize a minimal construction. 
Furthermore, the presence of the leptoquark leads to rich phenomenology, including semi-leptonic decays, neutral meson mixing, lepton flavor violations and lepton $g-2$, thereby rendering the model experimentally testable. 
After formulating each sector of our model, we perform a comprehensive numerical analysis, taking into account all relevant experimental constraints for both normal and inverted hierarchies of neutrino masses. 
Our analysis reveals characteristic tendencies within the viable parameter space. 
}
\end{abstract}
\maketitle
\newpage

\section{Introduction}

Elucidating the origin of neutrino masses remains one of the most important challenges in particle physics and strongly suggests the existence of physics beyond the Standard Model (BSM). 
Owing to their electrically neutral nature, neutrinos can be Majorana particles. The simplest way to generate Majorana neutrino masses is to introduce heavy Majorana right-handed neutrinos $N_R$. 
The relevant Lagrangian is given by
\begin{align}
- \mathcal{L} \supset y_D \overline{L_L} \tilde{H} N_R + M_R \overline{N_R^C} N_R + {\rm h.c.} \, ,
\end{align}
where $L_L \equiv [\nu_L, \ell_L]^T$ and $H \equiv [w^+, (v_h + h + iz) / \sqrt{2}]^T$ denote the isospin doublet leptons and Higgs boson in the SM, respectively, and $\tilde{H} \equiv i \tau_2 H^*$. 
Here $\tau_2$ is the second Pauli matrix, and $v_h \approx 246 \, {\rm GeV}$ is the vacuum expectation value (VEV) of the SM Higgs. 
The fields $w^+$ and $z$ are absorbed by charged-gauge boson $W^+$ and the neutral gauge boson $Z$, respectively, to provide their masses. 
After electroweak symmetry breaking, the active neutrino mass matrix is given by
\begin{align}
m_{\nu} \simeq - m_D M_R^{-1} m_D^T + {\rm h.c.} \, ,
\label{eq:dirac}
\end{align}
when we assume $m_D (\equiv y_D v_h / \sqrt{2}) \ll M_R$. 

Let us set $m_{\nu} \sim 100 \, {\rm meV}$ and $M_R \sim 100 \, {\rm TeV}$, where each value is motivated by the upper limit inferred from the minimal standard cosmological model and the maximal energy scale accessible at current or near-future experiments, such as CEPC-SPPC~\cite{CEPCStudyGroup:2023quu}, respectively. 
According to Eq.~\eqref{eq:dirac}, the Dirac Yukawa coupling $y_D$ is estimated as
\begin{align}
y_D \approx 1.82 \times 10^{-5} \, .
\end{align}
For comparison, the electron and muon Yukawa couplings in the SM are of the order $10^{-6}$ and $10^{-4}$, respectively. 
Such small values suggest that the possible existence of new physics responsible for generating these Yukawa structures. 
Indeed, it is well-known that these small Yukawa couplings can be generated radiatively~\cite{Okada:2013iba,Okada:2014nsa,Nomura:2016emz,Nomura:2017ezy,Nomura:2025sod,Nomura:2025tvz,Chen:2025awz} or through lepton seesaw mechanisms~\cite{Lee:2021gnw,Nomura:2025bph,Nomura:2024ctl} to alleviate their hierarchical structure. 
Relaxed Yukawa hierarchies, i.e., comparatively larger Yukawa couplings, often lead to rich phenomenology such as lepton flavor violations (LFVs), lepton anomalous magnetic dipole moment $\Delta a_\ell$ (lepton $g-2$), electric dipole moments (EDMs), and dark matter candidate. 
In this direction of model building, two guiding principles are particularly important: {\it minimal particle content (minimality)} and {\it experimental testability}. 
{\it Minimality} generally favors the introduction of bosons rather than additional fermions, since at least two generations of new fermions are required to reproduce neutrino oscillation data~\cite{Esteban:2024eli}. 
{\it Experimental testability} motivates the introduction of leptoquark boson, denoted by $S$ in this work; it is $SU(2)$ singlet with hypercharge $1/3$ that is commonly denoted as $S_1$~\cite{Dorsner:2016wpm}. 
Since $S$ couples simultaneously to leptons and quarks, leptoquark models are generically constrained by semi-leptonic decay processes and neutral meson mixing, in addition to LFVs and lepton $g-2$~\cite{Dorsner:2016wpm, AristizabalSierra:2007nf, Cheung:2016fjo, Chen:2017hir, Chen:2016dip}.\footnote{Moreover, in some models, no additional symmetries are required, since the specific electric charges of leptoquarks automatically forbid unwanted interactions.} 
From these perspectives, {\it leptoquarks provide one of the most promising candidates that simultaneously satisfy minimal particle content and experimental testability.} 
On the other hand, one must avoid excessively rapid nucleon decays mediated by leptoquarks. 
Note that current collider experiments, such as CERN Large Hadron Collider (LHC), constrain the leptoquark mass to be above $1 \, {\rm TeV}$. 

In our paper, a minimal one-loop induced Dirac neutrino mass model is proposed, in order to relax the Yukawa hierarchies. 
We introduce one scalar leptoquark $S$ with hypercharge $1/3$ transforming as $\bar{\bf 3}$ under the SM color $SU(3)_C$, while singlet under $SU(2)_L$. 
To forbid the tree-level Dirac mass term $\overline{L_L} \tilde H N_R$, we impose the Ising Fusion Rule (IFR) whose multiplication rules, consisting of three mutually commuting generators ($\mathbbm{I}, \sigma, \epsilon$), are simply given by~\cite{Kobayashi:2025cwx}
\begin{align}
\epsilon \otimes \epsilon = \mathbbm{I} \, , \quad 
\sigma \otimes \sigma = \mathbbm{I} \oplus \epsilon \, , \quad
\sigma \otimes \epsilon = \sigma \, .
\end{align}
This is the minimal symmetry structure capable of realizing our framework, and a Dirac neutrino mass matrix is then generated at one-loop level after dynamical breaking of the IFR, by appropriately assigning these generators into relevant particles. 
It is worth emphasizing that loop-level dynamical breaking is a distinctive feature of non-invertible symmetries\footnote{Relevant references can be found in Ref.~\cite{Kashav:2026jjg,Heckman:2024obe,Kaidi:2024wio,Suzuki:2025oov,Kobayashi:2025cwx,Nomura:2025sod,Chen:2025awz,Okada:2025kfm,Jangid:2025krp,Jangid:2025thp,Nomura:2025tvz,Okada:2025adm,Okada:2026gxl}.}, and cannot be realized within conventional group symmetries such as discrete Abelian $Z_N$ and/or non-Abelian discrete symmetries~\cite{Ishimori:2010au,Altarelli:2010gt,Kobayashi:2022moq}. 
As a consequence, the active neutrino masses arise effectively at two-loop level. 
Importantly, the IFR also eliminates dangerous operators that would otherwise induce rapid nucleon decays. 
We further discuss the relevant phenomenological implications associated with the leptoquark $S$ and perform a numerical analysis to explore the viable parameter space under current experimental constraints such as neutrino oscillation data, LFVs, lepton $g-2$, semi-leptonic decays and neutral meson mixings. 

This paper is organized as follows. 
In section~\ref{sec:II}, we first show our model with appropriate charge assignments of the IFR for all particles and explain how we can obtain active neutrino masses through loop-induced Dirac-type neutrino Yukawa couplings. 
In addition, we explain all relevant constraints to our analysis both from quark and lepton sectors. 
In section~\ref{sec:III}, we demonstrate numerical analysis and find a viable and interesting parameter space in our model. 
Furthermore, we discuss representative features for both cases of normal hierarchy (NH) and inverted hierarchy (IH) for the active neutrino masses. 
Finally, we summarize and conclude in section~\ref{sec:IV}.

\section{Model setup}
\label{sec:II}

\begin{table}[t!]
\begin{tabular}{|c||c|c|c|c|c|c||c|c|}\hline\hline 
 & ~$Q_L$~ & ~$u_R$~ & ~$d_R$~ & ~$L_L$~ & ~${\ell_R}$~ & ~${N_R}$~ & ~$H$~ & ~{$S$}~ \\\hline\hline
$SU(3)_C$ & $\bm{3}$ & $\bm{3}$ & $\bm{3}$ & $\bm{1}$ & $\bm{1}$ & $\bm{1}$ & $\bm{1}$ & $\bar{\bm{3}}$ \\\hline
$SU(2)_L$ & $\bm{2}$ & $\bm{1}$ & $\bm{1}$ & $\bm{2}$ & $\bm{1}$ & $\bm{1}$& $\bm{2}$ & $\bm{1}$ \\\hline
$U(1)_Y$ & $\frac{1}{6}$ & $\frac{2}{3}$ & $-\frac{1}{3}$ & $-\frac{1}{2}$ & $-1$ & $0$ & $\frac{1}{2}$ & $\frac{1}{3}$ \\\hline
${\rm IFR}$ & $\sigma$ & $\sigma$ & $\sigma$ & $\mathbb{I}$ & $\mathbb{I}$ & $ \epsilon$ & $\mathbb{I}$ & $\sigma$ \\\hline
\end{tabular}
\caption{Charge assignments of relevant fermions and bosons
under $SU(3)_c\otimes SU(2)_L\otimes U(1)_Y \otimes {\rm IFR}$. 
All fermions have to have three generations. }
\label{tab:1}
\end{table}
Our particle contents and its charge assignments are shown in Table~\ref{tab:1}. 
With these charge assignments, the renormalizable Lagrangian for the SM fermions is given by
\begin{align}
- {\cal L} &\supset y^u_{ij} \overline{Q_{L_i}} \tilde{H} u_{R_j} + y^d_{ii} \overline{Q_{L_i}} H d_{R_i} + y^{\ell}_{aa} \overline{L_{L_a}} H \ell_{R_a} + {\rm h.c.} \, .
\label{eq:SM}
\end{align}
Among the three Yukawa matrices, it is possible to diagonalize two of them, without loss of generality. 
For the present analysis, we adopt $y^d$ and $y^{\ell}$ as the diagonal couplings. 
After the spontaneous symmetry breaking, the mass matrices for quark sector are given by $m'_u = y^u v_h / \sqrt{2}$ and $m_d = y^d v_h / \sqrt{2} = {\rm diag} [D_d, D_s, D_b]$. 
The mass matrix $m'_u$ is diagonalized by $m_u = V_{u_L}^{\dag} m'_u V_{u_R} = {\rm diag} [D_u, D_c, D_t]$, and therefore, the Cabibbo-Kobayashi-Maskawa (CKM) matrix $V_{\rm CKM}$ is identical to $V_{u_L}^{\dag}$. 
The charged-lepton masses are given by $m_{\ell} \equiv y^{\ell} v_h / \sqrt{2} = {\rm diag} [D_e, D_{\mu}, D_{\tau}]$. 
Similar to the quark sector, the observed mixing matrix of lepton sector is identical to be mixing matrix to diagonalize the neutrino mass matrix, as shown later. 

In addition to Eq.~\eqref{eq:SM}, the relevant Lagrangian including new particles is found as
\begin{align}
- {\cal L}^{\rm new} &= f_{ia} \overline{Q^C_{L_i}} L_{L_a} S + h_{ia} \overline{u^C_{R_i}} \ell_{R_a} S + g_{i\alpha}^{\dag} \overline{d^C_{R_i}} N_{R_\alpha} S + M_{R_{\alpha\alpha}} \overline{N^C_{R_\alpha}} N_{R_\alpha} + {\rm h.c.} \, ,
\label{eq:new}
\end{align}
where $M_R$ is also diagonal without loss of generality. 
We suppose $i, j, k, \cdots$ to be quark flavor indices, $a, b, c, \cdots$ to be charged-lepton flavor ones, and $\alpha, \beta, \gamma, \cdots$ to be the right-handed flavor ones. 
Notice that $\overline{L_L}N_R \tilde{H}$ term is forbidden by IFR. 
Among the Yukawa interactions in Eq.~\eqref{eq:new}, the first and third terms induce Dirac-type neutrino mass at one loop level as shown below. 
In the analysis of this work, we focus on these interactions and the second term associated with coupling $h_{ia}$ is not considered assuming the coupling constants are sufficiently small to avoid relevant phenomenological constraints. 

The scalar potential for $H$ and $S$ is simply given by
\begin{align}
{\cal V} &= - \mu_H^2 |H|^2 - \mu_S^2|S|^2 + \lambda_H |H|^4 + \lambda_S |S|^4 + \lambda_{H S} |H|^2 |S|^2 \, .
\label{eq:vp}
\end{align}
Inserting tad-pole condition $(\partial{\cal V}/\partial h)|_{h\to0}=0$, each mass of the SM Higgs and the leptoquark $S$ is given by
\begin{align} 
m_h^2 = 2 \lambda_H v_h^2 \, , \quad m_S^2 = \modYS{-} \mu_S^2 + \frac{\lambda_{HS} v^2_h}{2} \, .
\end{align}
In the numerical analysis, we take $m_S$ as a free parameter, and theoretical constraint from the bounded from below for the scalar potential is satisfied by choosing $\mu_S$ appropriately so that $\lambda_{HS} \leq \mathcal{O} (1)$.

\subsection{Active neutrino mass matrix}

\begin{figure}
\centering
\begin{tikzpicture}
\begin{feynman}
\vertex[label=left:\(L_L \, (\mathbbm{I})\)] (a) at (0,0);
\vertex[label=right:\(N_R \, (\epsilon)\)] (b) at (6,0);
\vertex (c) at (1,0);
\vertex[label=above:\(y^d\)] (m) at (3,0);
\vertex[label=above:\(Q_L \, (\sigma)\)] (l) at (2,0);
\vertex[label=above:\(d_R \, (\sigma)\)] (r) at (3.9,0);
\vertex (d) at (5,0);
\vertex[label=above:\(S\,(\sigma)\)] (e) at (3,2);
\diagram* {
(a) -- (c) -- [insertion={[size=2.5pt]0.5}] (d) -- (b),
(c) -- [scalar, quarter left] (e) -- [scalar, quarter left] (d),
};
\end{feynman}
\end{tikzpicture}
\caption{Dirac neutrino mass matrix at one-loop level after dynamical breaking of IFR. 
$\mathbbm{I}, \sigma, \epsilon$ in parentheses indicate corresponding assignments of our IFR. }
\label{fig:NuMass}
\end{figure}
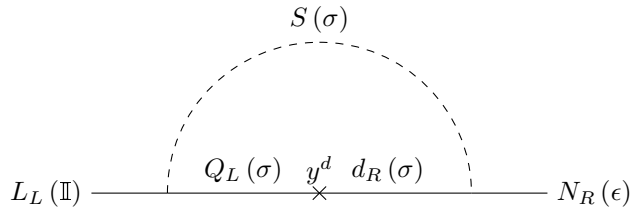

The Dirac neutrino mass matrix in Eq.~(\ref{eq:dirac}) is given by
\begin{align}
m_D \approx N_C \frac{g_{\alpha i} m_{d_i} f_{ia}} {(4 \pi)^2} \ln(r_S) \, , \label{eq:1lpd}
\end{align}
where we assume $m_d \ll m_S$, $N_C \equiv 3$ is the color factor, $r_S \equiv \frac{m_S^2}{\Lambda^2}$, and corresponding diagram is shown in Fig.~\ref{fig:NuMass}. 
%%%%%%%%%%%%%%%%%%%%%%%%%%%
%%%%%%%%%%%%%%%%%%%%%%%%%%%
Note here that we introduce the cut-off scale $\Lambda$ to regularize momentum integration in loop diagram as a minimal treatment; a divergence in radiative correction associated with dynamical breaking of non-invertible symmetry often appears, which should be regularized in some ways. 
In our numerical analysis, we fix it to be $100 \, {\rm TeV}$ which corresponds to the highest energy scale accessible to current or near-future experiments. 
Our choice of cut-off scale does not much affect phenomenology, since it only modifies the overall factor of $m_D$ at most $\mathcal{O}(10)$ level even if we choose $\Lambda$ as Planck scale instead of $100 \, {\rm TeV}$. 

The active neutrino mass matrix is given by Eq.~\eqref{eq:dirac} via type-I seesaw mechanism. 
When we define $m_\nu$ is diagonalized by $D_\nu = U^T_{\rm PMNS} m_\nu U_{\rm PMNS}$ with observed lepton mixing matrix $U_{\rm PMNS}$, $m_D$ can be rewritten in terms of the Casas-Ibarra parametrization~\cite{Casas:2001sr}
\begin{align}
m_D = i M^{1/2}_R O_N D^{1/2}_{\nu} U^{\dag}_{\rm PMNS} \, ,
\label{eq:ci}
\end{align}
where $O_N$ is a $3 \times 3$ orthogonal mixing matrix with three complex values. 
$U_{\rm PMNS}$ is given in Nufit~6.1~\cite{Esteban:2024eli}.\footnote{See \href{http://www.nu-fit.org/}{http://www.nu-fit.org/} for latest results.} 
Combining Eqs.~\eqref{eq:1lpd} and \eqref{eq:ci}, Yukawa coupling $g(f)$ is given by the following form:
\begin{align}
g &= i \left( \frac{N_C \ln(r_S)}{(4 \pi)^2} \right)^{-1} M^{1/2}_R O_N D^{1/2}_{\nu} U^{\dag} f^{-1} m_d^{-1} \, , \label{eq:g} \\
f &= i \left( \frac{N_C \ln(r_S)}{(4 \pi)^2} \right)^{-1} m_d^{-1} g^{-1} M^{1/2}_R O_N D^{1/2}_{\nu} U^{\dag}_{\rm PMNS} \, , \label{eq:f}
\end{align}
where we omit generation indices, and perturbation limit has to be imposed by $\max \bigl( |f(g)| \bigr) \le \sqrt{4 \pi}$. 
For convenience, we adopt Eq.~\eqref{eq:g} for our numerical analysis. 

In the neutrino sector, we have several observables and constraints on these. 
We will list all relevant constraints and future sensitivities to our discussion:\\
\noindent
\underline{\it The effective mass for neutrinoless double beta decay}, denoted by $m_{ee}$, is given by
\begin{align}
m_{ee} = \left| D_{\nu_1} \cos^2 \theta_{12} \cos^2 \theta_{13} + D_{\nu_2} \sin^2 \theta_{12} \cos^2 \theta_{13} e^{i \alpha_{2}} + D_{\nu_3} \sin^2 \theta_{13} e^{i (\alpha_{3} - 2 \delta_{CP})} \right| \, ,
\end{align}
where $\theta_{12, 23, 13}$ are mixing angles for $U_{\rm PMNS}$ in the standard parametrization, $\delta_{CP}$ is a Dirac CP phase, and $\alpha_{2, 3}$ are Majorana phases given by $P_{\rm Maj.} \equiv {\rm diag} (1, e^{\alpha_2/2}, e^{\alpha_3/2})$. 
The KamLAND-Zen collaboration provides an upper bound on $m_{ee}$ at the 90\% confidence level (CL)~\cite{KamLAND-Zen:2024eml}
\begin{align}
m_{ee} < (28-122) \, {\rm meV} \, .
\label{eq:mee_current}
\end{align}
Future experiments provide us more stringent constraints:
\begin{align}
m_{ee} < \begin{cases}
(9-21) \, {\rm meV} & \text{from LEGEND-1000~\cite{LEGEND:2021bnm}} \, , \\[0.3ex]
(4.7-20.3) \, {\rm meV} & \text{from nEXO~\cite{nEXO:2021ujk}} \, .
\end{cases}
\label{eq:mee_future}
\end{align}\\

\noindent
\underline{\it Direct search for neutrino mass}, denoted by $m_{\nu_e}^2 \equiv \sum_{i} D_{\nu_i}^2 |(U_{\rm PMNS})_{ei}|^2$, is known as model independent observable. 
Its explicit form is given by
\begin{align}
m_{\nu_e}^2 = \Bigl[ (D_{\nu_1} \cos \theta_{13} \cos \theta_{12})^2 + (D_{\nu_2} \cos \theta_{13} \sin \theta_{12})^2 + (D_{\nu_3} \sin \theta_{13})^2 \Bigr]^2 \, .
\end{align}
Experimental upper bound at 90\% CL from KATRIN~\cite{KATRIN:2024cdt} is given by
\begin{align}
m_{\nu_e} \le 450 \, {\rm meV} \, ,
\end{align}
which is weaker than the other constraints. \\

\noindent
\underline{\it The sum of neutrino masses}, denoted by $\sum D_{\nu} \equiv D_{\nu_1} + D_{\nu_2} + D_{\nu_3}$, is constrained by the minimal standard cosmological model and it provides the upper bound $\sum D_{\nu} \le 120 \, {\rm meV}$~\cite{Vagnozzi:2017ovm,Planck:2018vyg}.\footnote{Note here that it becomes weaker if the data are analyzed in the context of extended cosmological models~\cite{ParticleDataGroup:2014cgo}.
DESI and CMB data combination provides more stringent upper bound on the sum as $\sum D_{\nu} \le 72 \, {\rm meV}$~\cite{DESI:2024mwx}, although we do not apply the experiment for our analysis.} 
In numerical analysis, we will adopt NuFit~6.1~\cite{Esteban:2024eli} to experimental ranges of neutrino observables in addition to the above constraints of neutrino masses.

\subsection{LFVs and charged-lepton $g-2$}
\label{sec:pheno-l}

In this subsection, we will discuss LFVs and charged-lepton $g-2$. 
Each branching ratio for $\ell_a \to \ell_b \gamma$ denoted by ${\rm BR}(\ell_a \to \ell_b \gamma)$ and lepton $g-2$ denoted by $\Delta a_{\ell_a}$ are given by
\begin{align}
&{\rm BR} (\ell_a \to \ell_b \gamma) \approx \frac{48 \pi^3 \alpha_{\rm em}}{G_F^2 m^2_{\ell_a}} C_{ab} \left( \left| a_{L_{ab}} \right|^2 + \left| a_{R_{ab}} \right|^2 \right) \, , \label{eq:lfvs} \\
&\Delta a_{\ell_a} \approx - \frac{m_{\ell_a}}{2} \left(a_{L_{aa}} + a_{R_{aa}} \right) \, , \label{eq:da}
\end{align}
where $\ell_1 \equiv e$, $\ell_2 \equiv \mu$, $\ell_3 \equiv \tau$, $\alpha_{\rm em} \approx 1/137$ is the fine structure constant, and $G_F \approx 1.17 \times 10^{-5} \, {\rm GeV}^{-2}$ is the Fermi constant. 
The constant $C_{ab}$ depends on the process, and explicit values are $C_{21} \approx 1$, $C_{31} \approx 0.1784$, $C_{32} \approx 0.1736$, otherwise $C_{ab}$ is zero. 
$a_{L/R}$ depends on model, and our forms are found as
\begin{align}
a_{L_{ab}} &= \frac{m_{\ell_b}}{(4 \pi)^2} \sum_{i = 1}^3 f'^{\dag}_{bi} f'_{ia} \int[dx]_3 y \left[ \frac{z}{(x + y) m_S^2 + z m^2_{u_i} - x z m^2_{\ell_a} - y z m^2_{\ell_b}} \right. \nn \\
&\hspace{13.5em} \left. - \frac{2 x}{x m_S^2 + (1 - x) m^2_{u_i} - x z m^2_{\ell_a} - x y m^2_{\ell_b}} \right] \\
&\approx - \frac{m_{\ell_b}}{4 (4 \pi)^2 m^2_S} \sum_{i = 1}^3 f'^{\dag}_{bi} f'_{ia} \, , \\[0.5ex]
a_{R_{ab}} &= \frac{m_{\ell_a}}{(4 \pi)^2} \sum_{i = 1}^3 f'^{\dag}_{bi} f'_{ia} \int[dx]_3 x z \left[ \frac{1}{(x + y) m_S^2 + z m^2_{u_i} - x z m^2_{\ell_a} - y z m^2_{\ell_b}} \right. \nn \\
&\hspace{13.5em} \left. - \frac{2}{x m_S^2 + (1 - x) m^2_{u_i} - x z m^2_{\ell_a} - x y m^2_{\ell_b}} \right] \\
&\approx - \frac{m_{\ell_a}}{4 (4 \pi)^2 m^2_S} \sum_{i = 1}^3 f'^{\dag}_{bi} f'_{ia} \, ,
\end{align}
where $f' \equiv V_{\rm CKM}^T f$, $u_1 \equiv u$, $u_2 \equiv c$, $u_3 \equiv t$, $[dx]_3 \equiv dx dy dz \delta (1 - x - y - z)$, and $m_{u_i}, m_{\ell_a} \ll m_S$ is assumed in the last equations. 
 
The current upper bounds on ${\rm BR} (\ell_a \to \ell_b \gamma)$ are given at 90\% CL as~\cite{MEGII:2025gzr, MEGII:2023ltw, BaBar:2009hkt, Belle:2021ysv}
\begin{align}
{\rm BR} (\mu \to e \gamma) < 1.5 \times 10^{-13} \, , \quad {\rm BR} (\tau \to e \gamma) < 3.3 \times 10^{-8} \, , \quad {\rm BR} (\tau \to \mu \gamma) < 4.2 \times 10^{-8} \, .
\end{align}
The lepton $g-2$ receives the following experimental measurements~\cite{Fan:2022oyb,Fan:2022eto,ParticleDataGroup:2024cfk,Morel:2020dww}:
\begin{align}
\Delta a_e = (3.41 \pm 1.64) \times 10^{-13} \, , \quad \Delta a_{\mu} = (39 \pm 64) \times 10^{-11} \, .
\label{eq:g-2ell}
\end{align}
Here, $\Delta a_{\tau}$ is less precise compared with $\Delta a_{e, \mu}$, and therefore, we neglect our prediction for $\Delta a_{\tau}$ in the numerical analysis. 
In fact, since there is no specific mechanism to enhance $\Delta a_{\tau}$, its prediction is easily estimated via naive minimal flavor violation relation, $\Delta a_{\tau} / \Delta a_{\ell} \simeq ( m_{\tau} / m_{\ell} )^2$ for $\ell = e, \mu$.

\subsection{Semi-leptonic decays and neutral meson mixings}
\label{sec:pheno-q}

In this subsection, we will discuss semi-leptonic decays and neutral meson mixings. 
We start from effective Lagrangian with assuming $m_S$ is much larger than the other masses of SM fermions. 
The relevant 6-dimensional Lagrangian is given by
\begin{align}
{\cal L}^{\rm eff} &\sim \frac{f'_{i'a'} f^{\dag}_{ai}}{2 m_S^2} (\overline{d_i} \gamma_{\mu} P_L u_{i'}) (\overline{\nu_a} \gamma^{\mu} P_L \ell_{a'}) + \frac{f_{ia} f^{\dag}_{bj}}{2 m_S^2} (\overline{d_j} \gamma_{\mu} P_L d_{i}) (\overline{\nu_b} \gamma^{\mu} P_L \nu_{a}) \nn \\
&\hspace{1.2em} + \frac{f'_{i'a'} f'^{\dag}_{b'j'}}{2 m_S^2} (\overline{u_{j'}} \gamma_{\mu} P_L u_{i'}) (\overline{\ell_{b'}} \gamma^{\mu} P_L \ell_{a'}) \\
&\equiv \frac{1}{2} C^{ijkn}_{f'f} (\overline{\nu_i} \gamma^{\mu} P_L \ell_{j}) (\overline{d_k} \gamma_{\mu} P_L u_{n}) + \frac{1}{2} C^{ijkn}_{ff} (\overline{\nu_j}\gamma^{\mu} P_L \nu_{j}) (\overline{d_k} \gamma_{\mu} P_L d_{n}) \nn \\
&\hspace{1.2em} + \frac{1}{2} C^{ijkn}_{f'f'} (\overline{\ell_{i}}\gamma^{\mu} P_L \ell_{j}) (\overline{u_{k}} \gamma_{\mu} P_L u_{n}) \, ,
\end{align}
where we replace $(aa'ii') \to (ijkn)$ for $C^{ijkn}_{f'f}$, $(baji) \to (ijkn)$ for $C^{ijkn}_{ff}$ and $(b'a'j'i') \to (ijkn)$ for $C^{ijkn}_{f'f'}$. 
To check the experimental constraints for each process, it is convenient to define
\begin{align}
\epsilon^{ijkn}_{f^{(')} f^{(')}} \equiv \frac{\sqrt{2}}{4 G_F} C^{ijkn}_{f^{(')} f^{(')}} \, .
\end{align}
All constraints can be found in Ref.~\cite{Carpentier:2010ue}: Table 13 for $\epsilon^{ijkn}_{f'f}$, Table 12 for $\epsilon^{ijkn}_{ff}$ and Table 4 for $\epsilon^{ijkn}_{f'f'}$. 
For $\epsilon^{ijkn}_{f'f}$, there are lots of constraints, and some of them may be irrelevant to constrain our viable parameter space, due to flavor-blind processes. 
Therefore, instead, we conservatively impose their stringent bounds obtained as
\begin{align}
\max \left[ \epsilon^{ijkn}_{f'f} \right] \le 1.0 \times 10^{-3} \, .
\end{align}
Experimental upper bounds on $\epsilon^{ijkn}_{ff}$ can be borrowed as
\begin{align}
&\epsilon^{2211}_{ff} \le 7.3 \times 10^{-3} \, , \ \max \left[ \epsilon^{ij12}_{ff} \right] \le 9.4 \times 10^{-6} \, , \\
&\max \left[ \epsilon^{ij31}_{ff} \right] \le 4.9 \times 10^{-2} \, , \ \max \left[ \epsilon^{ij32}_{ff} \right] \le 1.0 \times 10^{-3} \, ,
\end{align}
while those on $\epsilon^{ijkn}_{f'f'}$ are
\begin{align}
&\epsilon^{1122}_{f'f'} < 1.0 \times 10^{-2} \, , \ \epsilon^{1133}_{f'f'} < 0.092 \, , \ \epsilon^{1222}_{f'f'} < 0.6 \, , \ \epsilon^{1322}_{f'f'} < 1.4 \, , \\
&-0.02 < \epsilon^{2211}_{f'f'} < 7.8 \times 10^{-3} \, , \ \epsilon^{2222}_{f'f'} < 0.4 \, , \ \epsilon^{2233}_{f'f'} < 0.061 \, , \ \epsilon^{2322}_{f'f'} < 1.6 \, , \\
&\epsilon^{3311}_{f'f'} < 0.54 \, , \ \epsilon^{3322}_{f'f'} < 0.54 \, , \ \epsilon^{3333}_{f'f'} < 0.086 \, .
\end{align}
Since most of them are flavor dependent constraints, we impose all of above bounds in the numerical analysis.\footnote{We do not consider the bound on $(0.01 <) \epsilon^{1111}_{f'f'}$, since the bound, which comes from Atomic Parity Violation, have a lot of experimental error ambiguity.} 

The neutral meson mixings are induced via box diagrams, and each form can be estimated as
\begin{align}
\Delta m_{M_d}^{ijk\ell} \simeq \frac{m_{M_d} f^2_{M_d}}{3 (4 \pi)^2 m^2_S} \sum_{\alpha, \beta = 1}^3 {\rm Re} &\left[ (f_{i \alpha} f^{\dag}_{\alpha j}) (f_{\ell \beta} f^{\dag}_{\beta k}) + (f_{i \alpha} f^{\dag}_{\alpha k}) (f_{\ell \beta} f^{\dag}_{\beta j}) \right. \nn \\
&\hspace{0.5em} + \left( (g^{\dag}_{i \alpha} g_{\alpha j}) (g^{\dag}_{\ell \beta} g_{\beta k}) + (g^{\dag}_{i \alpha} g_{\alpha k}) (g^{\dag}_{\ell \beta} g_{\beta j}) \right) G_1(r_{\alpha}, r_{\beta}) \nn \\
&\hspace{0.5em} \left. - \frac{1}{m_S^2} (g^{\dag}_{i \alpha} M_{R_{\alpha}} g^*_{\alpha \ell}) (g^T_{k \beta} M_{R_{\beta}} g_{\beta j}) G_2(r_{\alpha}, r_{\beta})
\right] \, , \\
\Delta m_{M_u}^{ijk\ell} \simeq \frac{m_{M_u} f^2_{M_u}}{3 (4 \pi)^2 m^2_S} \sum_{\alpha, \beta = 1}^3 {\rm Re} &\left[ (f'_{i \alpha} f'^{\dag}_{\alpha j}) (f'_{\ell \beta} f'^{\dag}_{\beta k}) + (f'_{i \alpha} f'^{\dag}_{\alpha k}) (f'_{\ell \beta} f'^{\dag}_{\beta j}) \right] \, .
\end{align}
where 
\begin{align}
G_1(r_{\alpha}, r_{\beta}) &\approx 2 \int [dx]_3 \frac{x}{x + y r_{\alpha} +z r_{\beta}} \, , \\
G_2(r_{\alpha}, r_{\beta}) &\approx \int [dx]_3 \frac{x}{(x + y r_{\alpha} +z r_{\beta})^2} \, .
\end{align}
Here, $m_{M_{u/d}}$ and $f_{M_{u/d}}$ are the mass and the decay constant of each meson ($M_u = D$ and $M_d = K, B_d, B_s$), respectively. 
The experimental results provide their upper bounds that are given by~\cite{Gabbiani:1996hi, ParticleDataGroup:2024cfk}:
\begin{align}
{\rm K^0(d\bar s)-\bar K^0(s\bar d)\ mixing} &:\ \Delta m_K \equiv \Delta m_K^{\rm SM} + \Delta m_K^{1221} \lesssim 3.484 (6) \times 10^{-15} \, {\rm GeV} \, , \label{eq:kk} \\
%%%%%%%%%%
{\rm B_d^0(d\bar b)-\bar B_d^0(b\bar d)\ mixing} &:\ \Delta m_{B_d} \equiv \Delta m_{B_d}^{\rm SM} + \Delta m_{B_d}^{1331} \lesssim 3.336 (13) \times 10^{-13} \, {\rm GeV} \, , \label{eq:bd} \\
%%%%%%%%%%
{\rm B_s^0(s\bar b)-\bar B_s^0(b\bar s)\ mixing} &:\ \Delta m_{B_s} \equiv \Delta m_{B_s}^{\rm SM} + \Delta m_{B_s}^{2332} \lesssim 1.1693 (4) \times 10^{-11} \, {\rm GeV} \, , \label{eq:bs} \\
% ref: p.111 "http://pdg.lbl.gov/2016/tables/rpp2016-sum-mesons.pdf"
%%%%%%%%%%%
{\rm D^0(c\bar u)-\bar D^0(u\bar c)\ mixing} &:\ \Delta m_D \equiv \Delta m_D^{\rm SM} + \Delta m_D^{2112} \lesssim 6.562 (764) \times 10^{-15} \, {\rm GeV} \, . \label{eq:dd} 
\end{align}
The following parameters are used in our analysis: $(f_K, m_K) \approx (0.156, 0.498) \, {\rm GeV}$, $(f_{B_d}, m_{B_d}) \approx (0.192, 5.280) \, {\rm GeV}$, $(f_{B_s}, m_{B_s}) \approx (0.274, 5.367) \, {\rm GeV}$, $(f_D, m_D) \approx (0.210, 1.865) \, {\rm GeV}$. 
Each SM prediction $\Delta m_{K, B_d, B_s}^{\rm SM}$ can be found as~\cite{Mescia:2012fg,Albrecht:2024oyn}
\begin{align}
\Delta m_K^{\rm SM} = 3.1 (1.2) \times 10^{-15} \, {\rm GeV} \, , \ \Delta m_{B_d}^{\rm SM} = 3.52 (14) \times 10^{-13} \, {\rm GeV} \, , \ \Delta m_{B_s}^{\rm SM} = 1.20 (4) \times 10^{-11} \, {\rm GeV} \, , \label{eq:DelmMSM}
\end{align}
while that of $D^0-\bar{D}^0$ mixing, $\Delta m_D^{\rm SM}$, involves large uncertainty due to non-perturbative effects (see, e.g., Ref.~\cite{Dulibic:2025emg} and references therein). 
Therefore, we neglect this contribution and set $\Delta m_D^{\rm SM} = 0$ in our analysis. 
Taking into account these predictions, we find the maximum allowed contributions from our model as follows:
\begin{align}
\Delta m_K^{1221} &< \Delta m_K^{\rm BSM, \ max} = 2.8 \times 10^{-15} \, {\rm GeV} \, , \\
\Delta m_{B_d}^{1331} &< \Delta m_{B_d}^{\rm BSM, \ max} = 0.097 \times 10^{-13} \, {\rm GeV} \, , \\
\Delta m_{B_s}^{2332} &< \Delta m_{B_s}^{\rm BSM, \ max} = 0.049 \times 10^{-11} \, {\rm GeV} \, , \\
\Delta m_D^{2112} &< \Delta m_D^{\rm BSM, \ max} = 6.562 \times 10^{-15} \, {\rm GeV} \, .
\end{align}
In this analysis, we incorporate 2$\sigma$ uncertainties from both experimental results and theoretical predictions.

\section{Numerical results}
\label{sec:III}

In this section, we perform the numerical analysis, taking into account all constraints discussed above. 
At first, we randomly select  our input parameters within the following ranges:
\begin{align}
&|\theta_1^O, \theta_2^O, \theta_3^O| = [10^{-5}, 1] \, , \quad [\alpha_2, \alpha_3] = [-\pi, \pi] \, , \quad |f| = [10^{-5}, \sqrt{4 \pi}] \, , \\
&D_{\nu}^{\rm lightest} = [10^{-13}, 10^{-10}] \, {\rm GeV} \, , \quad m_S = [10^3, 10^5] \, {\rm GeV} \, , \label{eq:neut-S} \\
&M_{R_1} = [10^2, 10^5] \, {\rm GeV} \, , ~ M_{R_2} = [M_{R_1}, 10^5] \, {\rm GeV} \, , ~ M_{R_3} = [M_{R_2}, 10^5] \, {\rm GeV} \, , 
\end{align}
where $\theta_1^O, \theta_2^O, \theta_3^O$ are mixing angles of $O_N$ in Eq.~\eqref{eq:ci}, $\alpha_{2, 3}$ are the Majorana phases, and $D_{\nu}^{\rm lightest}$ is $D_{\nu_1}$ for NH and $D_{\nu_3}$ for IH. 
As for the neutrino oscillation data, we apply them to the best fit values (BFs) of Nufit~6.1~\cite{Esteban:2024eli}. 
Within these ranges, we scan the parameter space to identify the region satisfying all the experimental constraints discussed above. 
In quark sector and the masses of charged-lepton, we refer BFs of experimental values from Particle Data Group~\cite{ParticleDataGroup:2024cfk}:
\begin{align}
&\sin \theta_{12}^Q = 0.22501 \, , \quad \sin \theta_{23}^Q = 0.04183 \, , \quad \sin \theta_{13}^Q = 0.003732 \, , \quad \delta_Q = 65.7183^{\circ} \, , \\
&[m_u, m_c, m_t] = [2.16 \times 10^{-3}, 1.2730, 172.56] \, {\rm GeV} \, , \\
&[m_d, m_s, m_b] = [4.70 \times 10^{-3}, 0.935, 4.183] \, {\rm GeV} \, , \\
&[m_e, m_{\mu}, m_{\tau}] = [0.511 \times 10^{-3}, 0.1057, 1.777] \, {\rm GeV} \, .
\end{align}

\subsection{NH}

In case of NH, we fix the BFs of the neutrino oscillation data as follows:
\begin{align}
&\sin \theta_{12}^2 = 0.3088 \, , \quad \sin \theta_{23}^2 = 0.470 \, , \quad \sin \theta_{13}^2 = 0.02248 \, , \quad \delta_{\rm CP} = 207^{\circ} \, , \nn \\
&\Delta m_{\rm sol}^2 = 7.537 \times 10^{-5} \, {\rm eV}^2 \, , \quad \Delta m_{\rm atm}^2 \equiv D_{\nu_3}^2 - D_{\nu_1}^2 = 2.521 \times 10^{-3} \, {\rm eV}^2 \, .
\end{align}
%%%%%%%%%%%%%%%%%%%
\begin{figure}[tb]\begin{center}
\includegraphics[width=80mm]{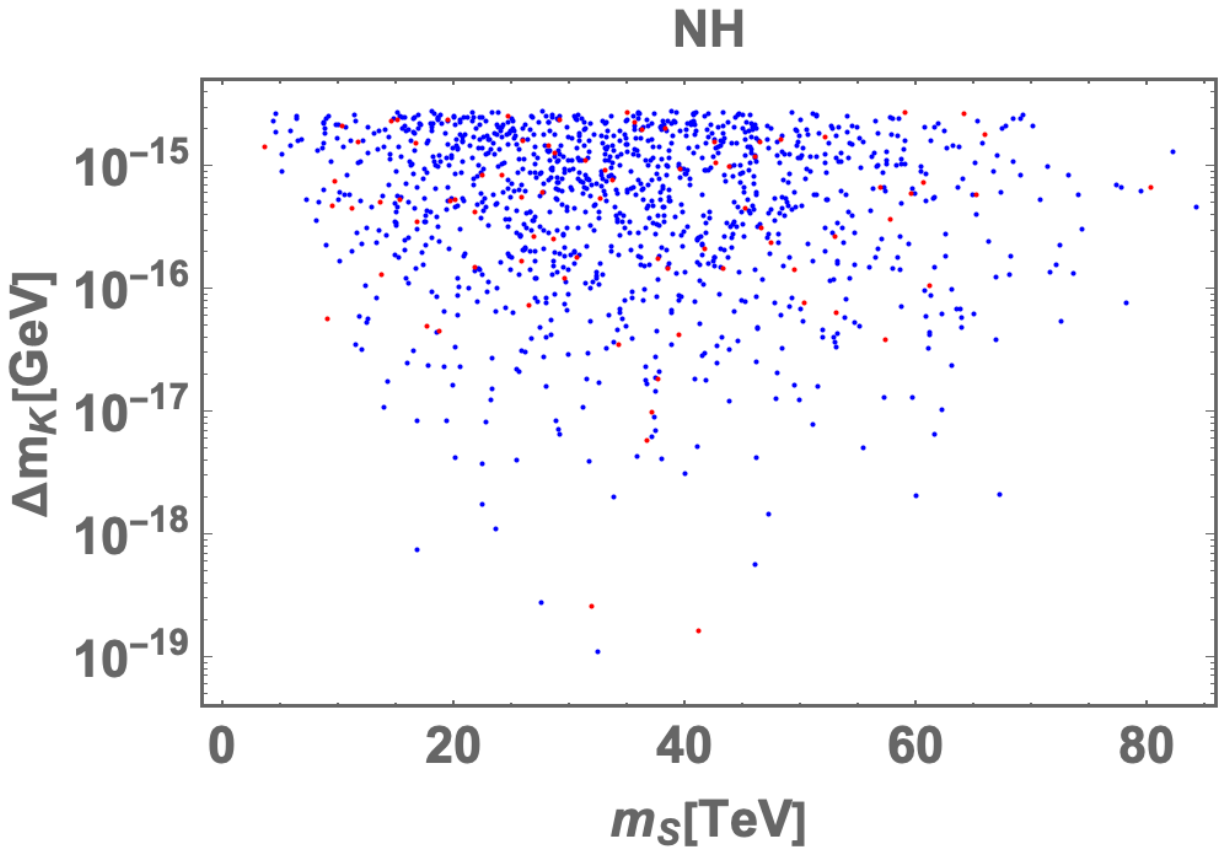}
\includegraphics[width=80mm]{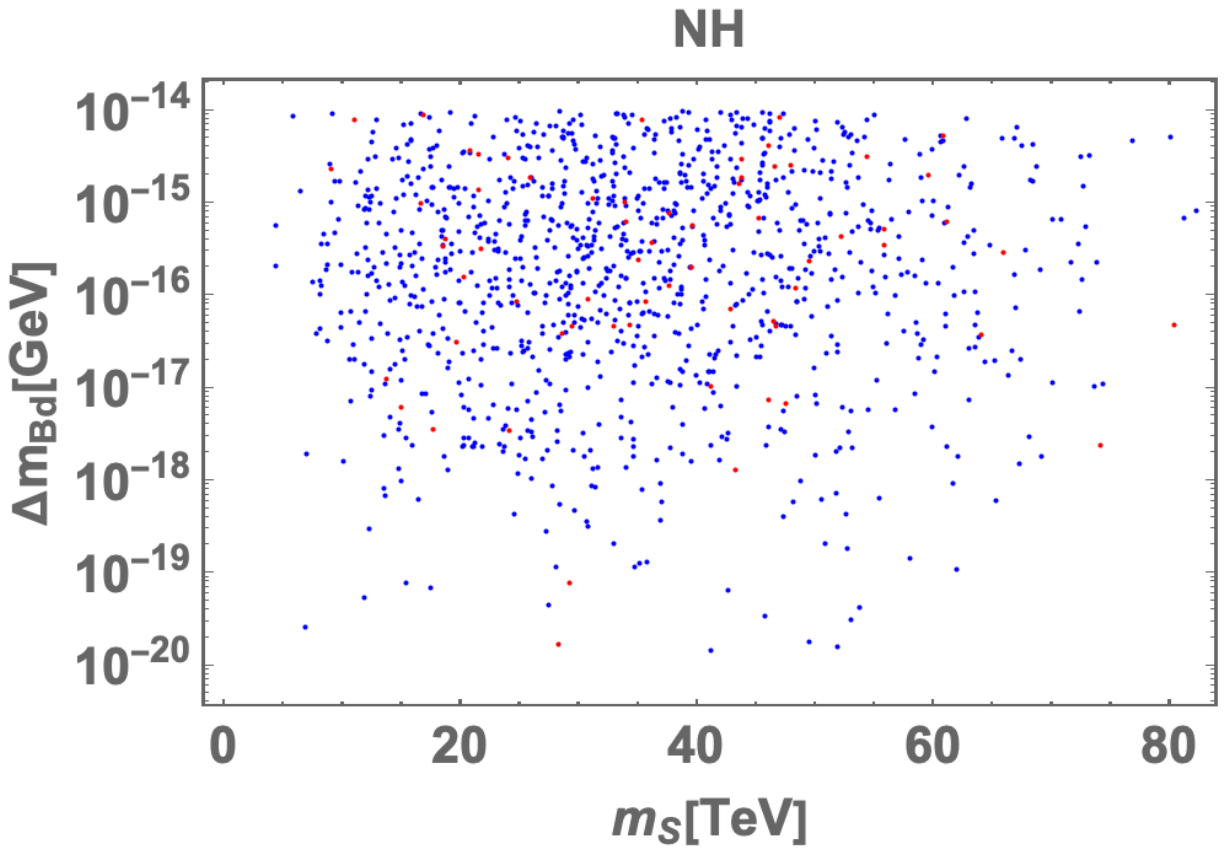}
\includegraphics[width=80mm]{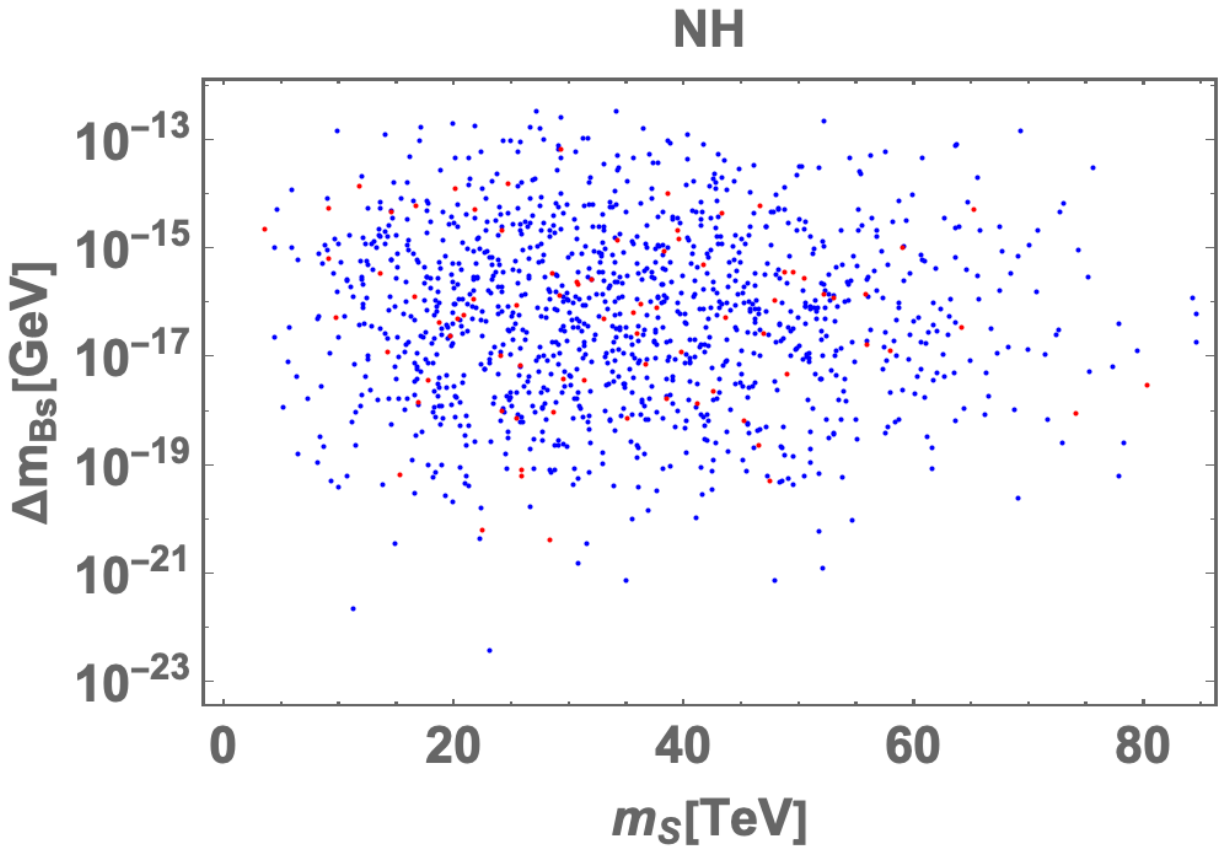}
\includegraphics[width=80mm]{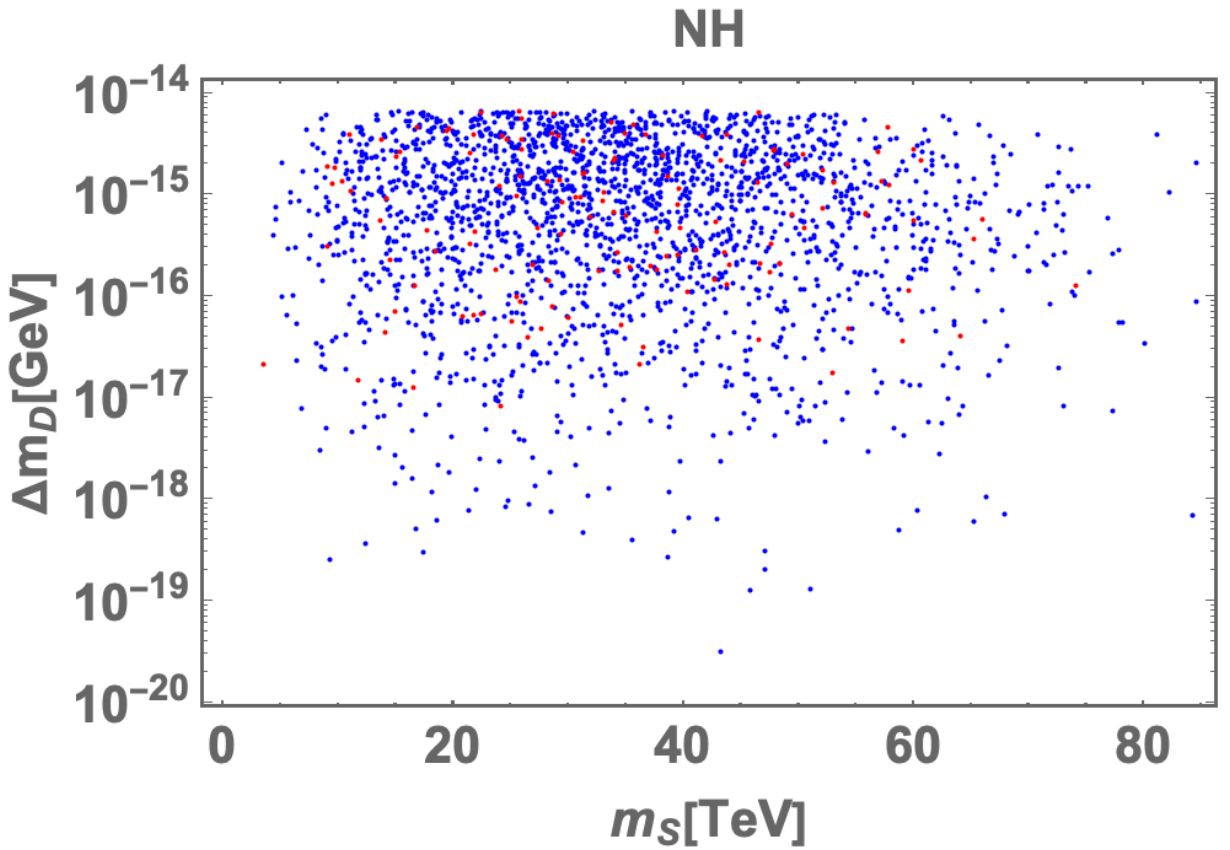}
\caption{Neutral meson mixings in terms of $m_S$ in case of NH. 
Here, blue points satisfy $\sum D_{\nu} \le 120 \, {\rm meV}$, while the red ones do not satisfy. }
\label{fig:nh1}
\end{center}\end{figure}
%%%%%%%%%%%%%%%%%%%
Using these values, our predictions of neutral meson mixings for $\Delta m_K$ (upper-left), $\Delta m_{B_d}$ (upper-right), $\Delta m_{B_s}$ (lower-left) and $\Delta m_D$ (lower-right) are shown in Fig.~\ref{fig:nh1}, in terms of $m_S$. 
The blue points satisfy $\sum D_{\nu} \le 120 \, {\rm meV}$, while the red ones do not satisfy. 
Hereafter, we use this color legends to all plots, without mentioning. 
These figures suggest that allowed regions of $\Delta m_{B_d}$ and $\Delta m_{B_s}$ tend to be less than each upper bound, while $\Delta m_K$ and $\Delta m_D$ tend to be located to nearby their upper bounds. 
It implies that $\Delta m_K$ and $\Delta m_D$ provide severe constraints to our model. 
In other words, our model can be tested by these observables, although the SM prediction of $\Delta m_D$ as well as its experimental sensitivity are not precisely obtained at present. 
Note that $m_S$ runs over wide range of $[4 \, {\rm TeV} \mathchar`- 85 \, {\rm TeV}]$ within the allowed region in Eq.~(\ref{eq:neut-S}). 

%%%%%%%%%%%%%%%%%%%
\begin{figure}[tb]\begin{center}
\includegraphics[width=80mm]{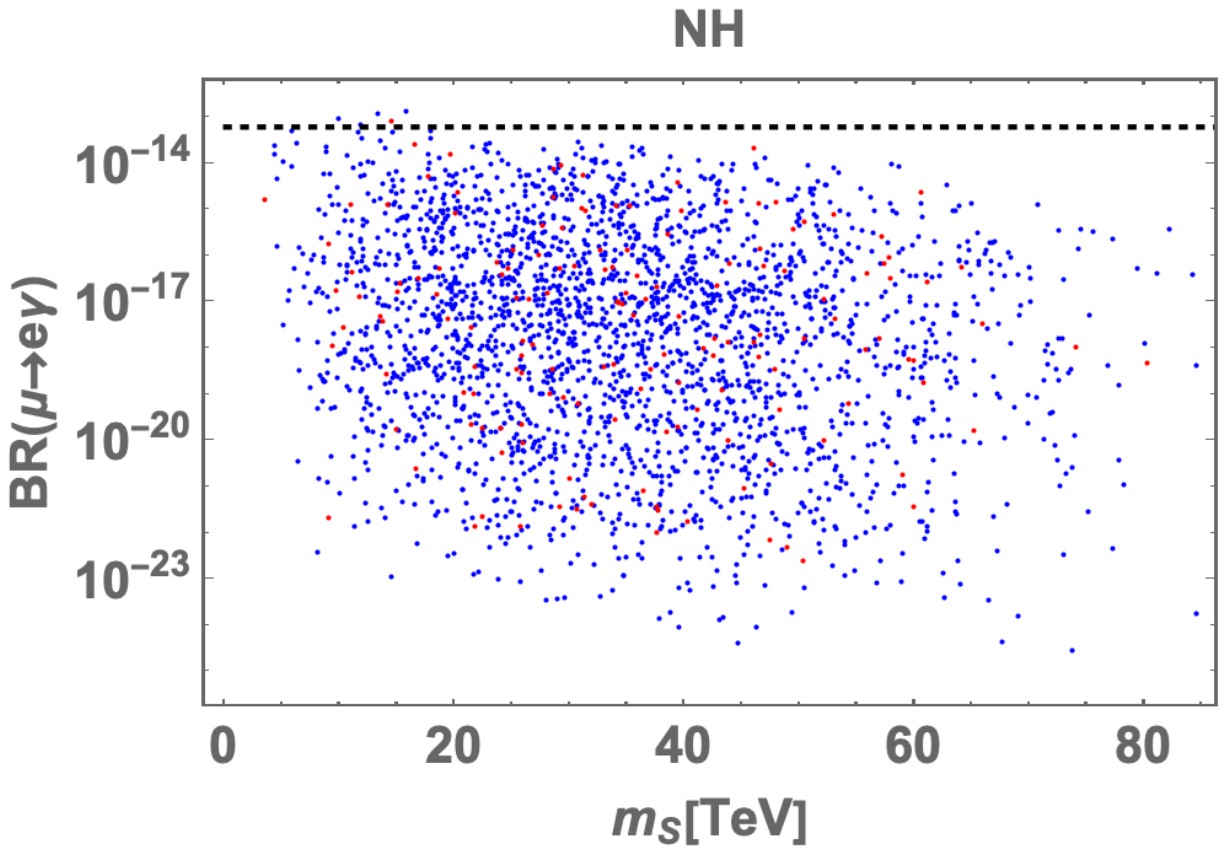}
\includegraphics[width=80mm]{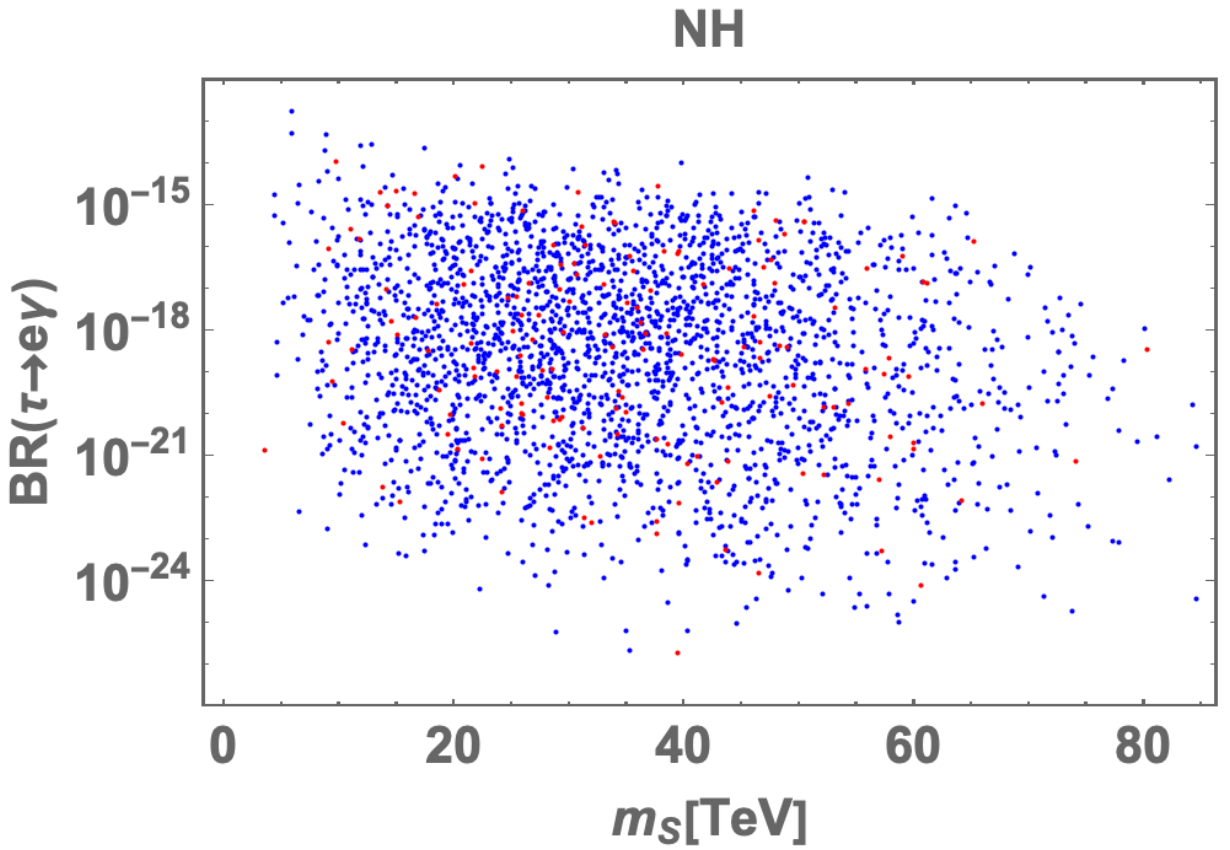}
\includegraphics[width=80mm]{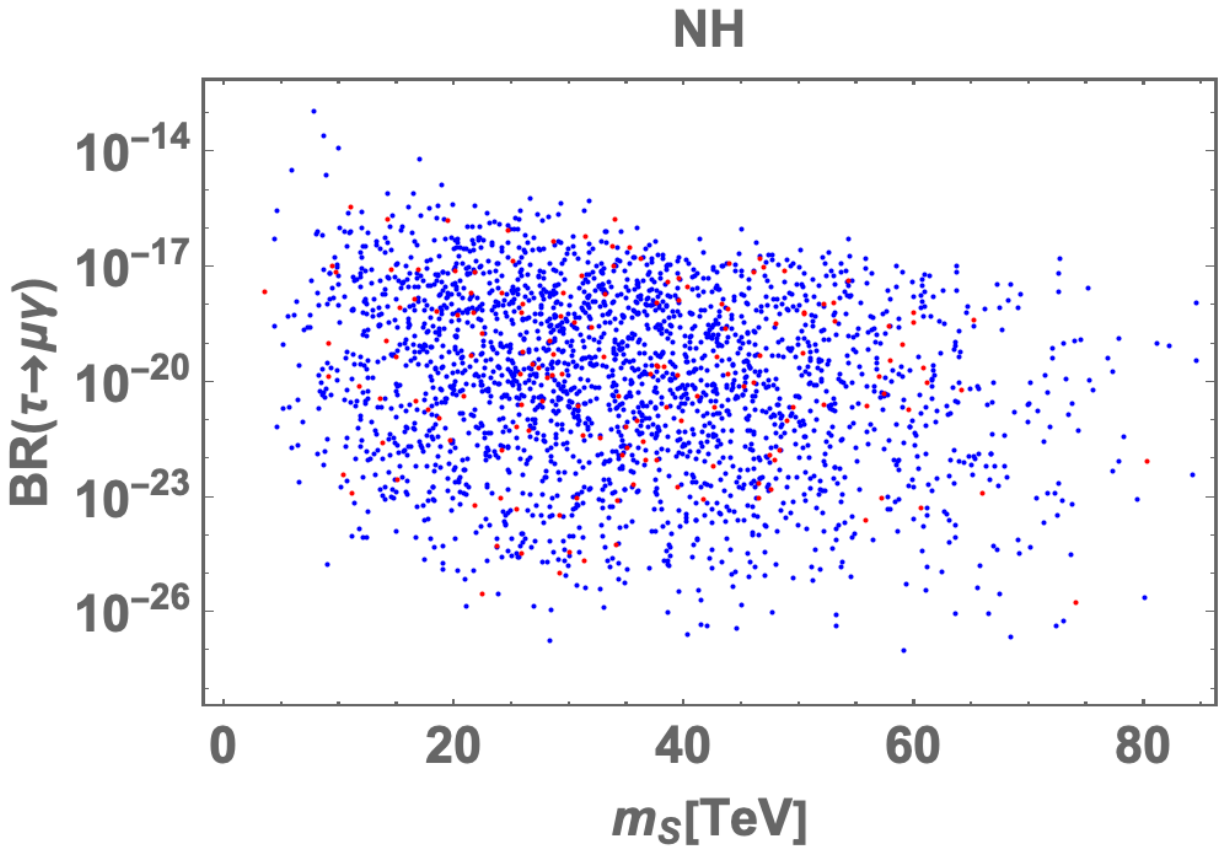}
\includegraphics[width=80mm]{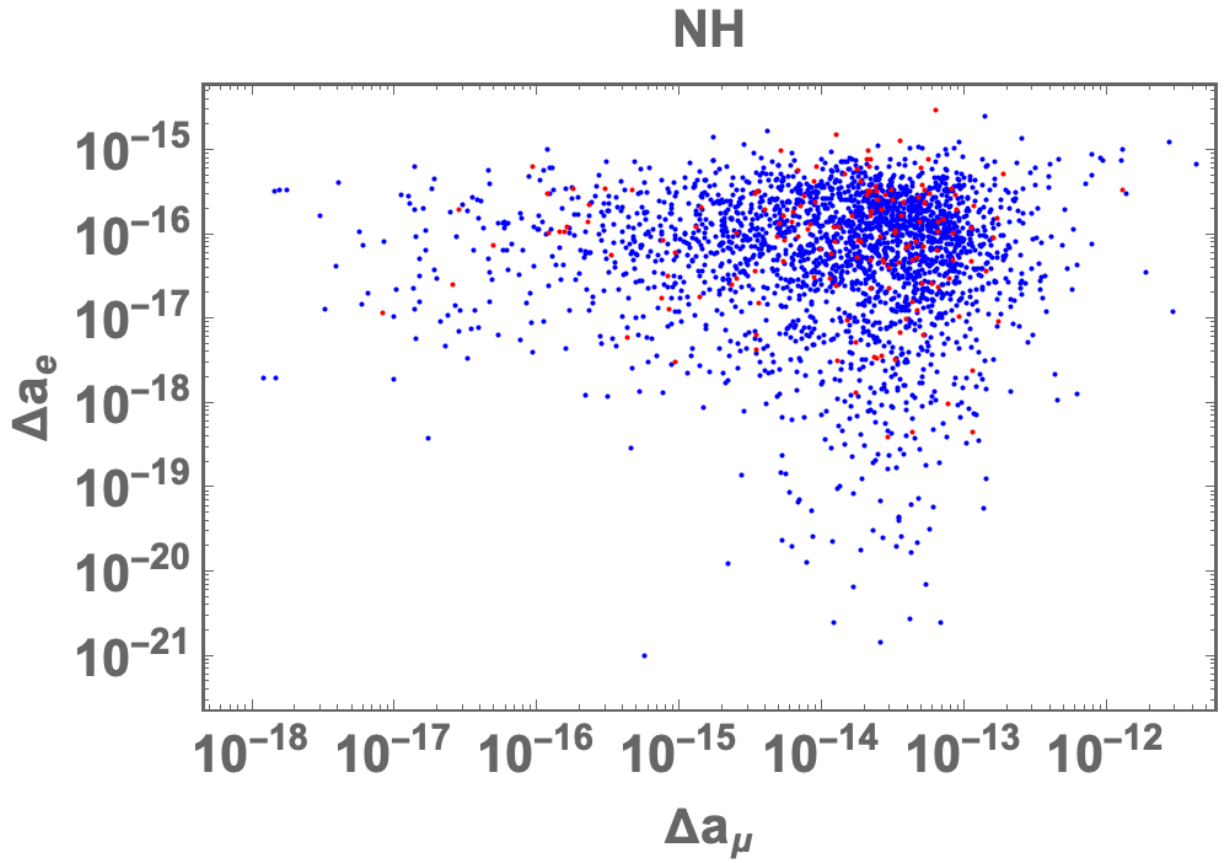}
\caption{Predictions of LFVs and lepton $g-2$ in case of NH, where the color legends are the same as the ones in Fig.~\ref{fig:nh1}. 
The dotted horizontal line of ${\rm BR} (\mu \to e \gamma)$ represents the futures accessible sensitivity, ${\rm BR} (\mu \to e \gamma) = 6 \times 10^{-14}$. }
\label{fig:nh2}
\end{center}\end{figure}
%%%%%%%%%%%%%%%%%%%
In Fig.~\ref{fig:nh2}, we show our predictions of branching ratios for three LFVs, ${\rm BR} (\mu \to e \gamma)$ (upper-left), ${\rm BR} (\tau \to e \gamma)$ (upper-right), and ${\rm BR} (\tau \to \mu \gamma)$ (lower-left) in terms of $m_S$, and $\Delta a_e$ versus $\Delta a_{\mu}$ (lower-right). 
The allowed regions in all LFVs are less than upper bound on $\mu \to e \gamma$. 
Thus, only the ${\rm BR} (\mu \to e \gamma)$ could come into a good testable observable near future, as one can see from the upper-left panel: for lighter region of $m_S = [10 \, {\rm TeV} \mathchar`- 16 \, {\rm TeV}]$, there are some allowed points above the dotted horizontal line which corresponds to the future accessible sensitivity, ${\rm BR} (\mu \to e \gamma) = 6 \times 10^{-14}$~\cite{MEGII:2018kmf}. 
In the lepton $g-2$, on the other hand, allowed regions tend to be localized at nearby $\Delta a_{\mu} = [10^{-14} \mathchar`- 10^{-13}]$ and $\Delta a_e = [10^{-17} \mathchar`- 10^{-15}]$ which are several orders of magnitude smaller than each result in Eq.~\eqref{eq:g-2ell}, although wide ranges are allowed. 

%%%%%%%%%%%%%%%%%%%
\begin{figure}[tb]\begin{center}
\includegraphics[width=80mm]{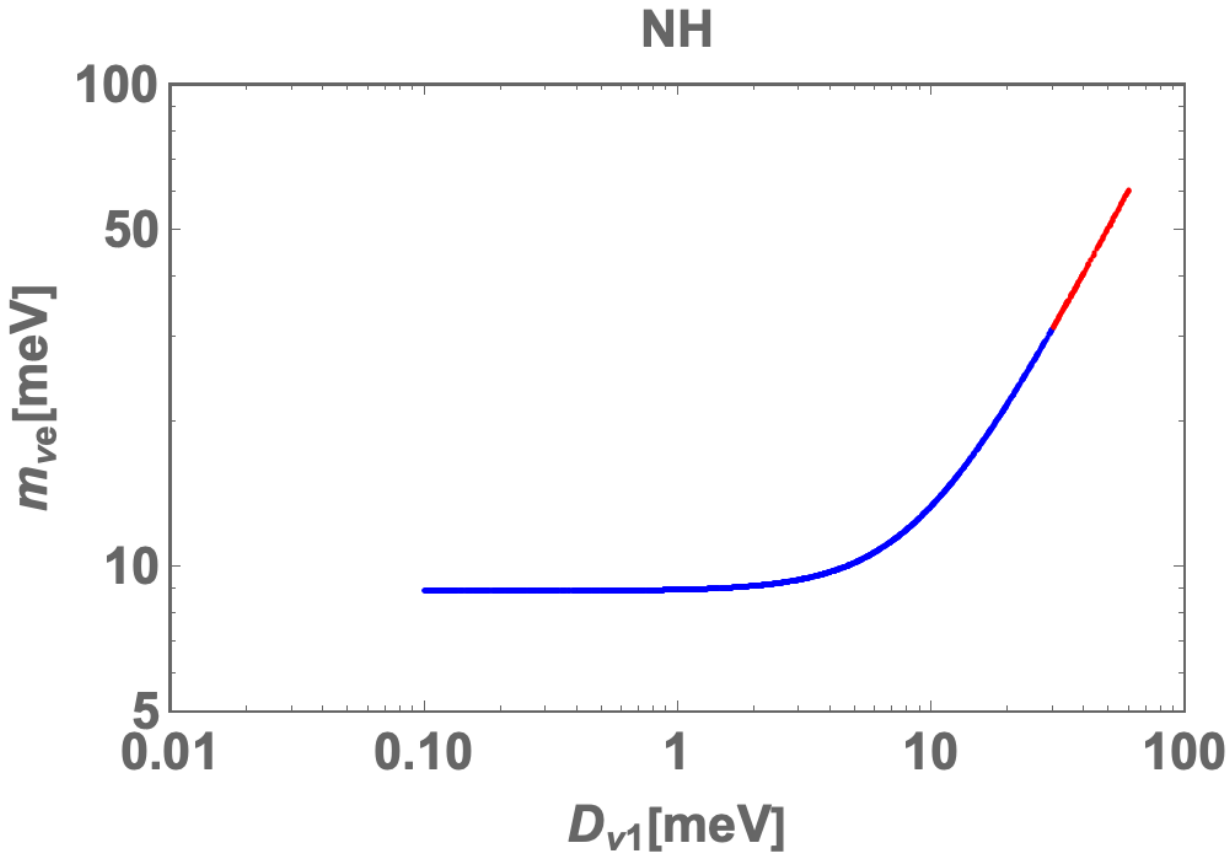}
\includegraphics[width=80mm]{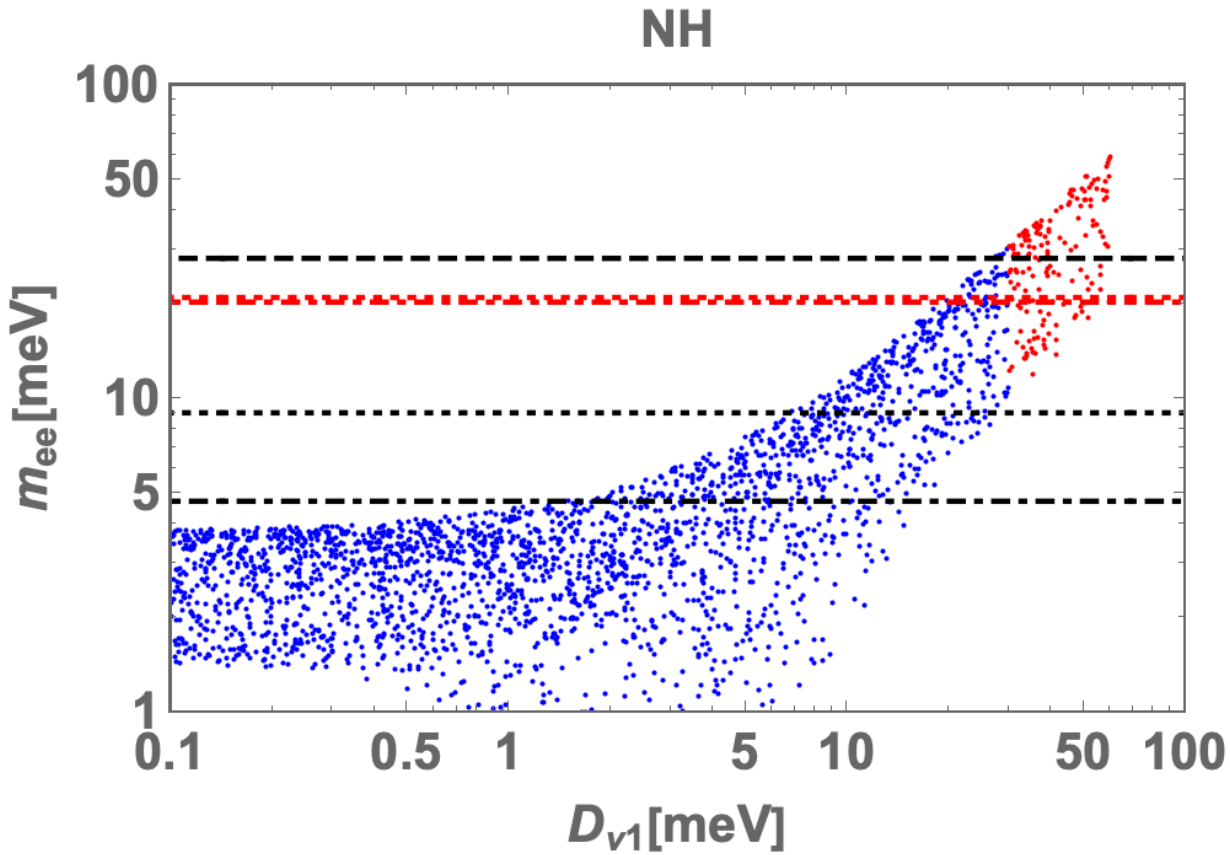}
\caption{Neutrino observations in case of NH. }
\label{fig:nh3}
\end{center}\end{figure}
%%%%%%%%%%%%%%%%%%%
Finally, we show $m_{\nu_e}$ (left) and $m_{ee}$ (right) in terms of the lightest neutrino mass $D_{\nu_1}$ in Fig.~\ref{fig:nh3}. 
The result for $m_{\nu_e}$ in the left panel tells us that the allowed region is in the range of $[9 \, {\rm meV} \mathchar`- 60 \, {\rm meV}]$, but that becomes narrower up to $30 \, {\rm meV}$ when $\sum D_{\nu} \le 120 \, {\rm meV}$ is considered. 
The range is too small to reach the current experimental bound, $m_{\nu_e} \leq 450 \, {\rm meV}$. 
On the other hand, for $m_{ee}$ result in right panel, the allowed region is reached the current experimental bounds, and some region can be tested by several future experiments: the black dashed horizontal line corresponds to the current lowest upper bound, $28 \, {\rm meV}$ from KamLAND-Zen, and this bound excludes the region $30 \, {\rm meV} \lesssim D_{\nu_1}$ in combination of $\sum D_{\nu} \le 120 \, {\rm meV}$. 
The LEGEND-1000 experiment has potential to explore our allowed region of $20 \, {\rm meV} \lesssim D_{\nu_1} \lesssim 30 \, {\rm meV}$ for the highest upper bound (red dashed line) or $7 \, {\rm meV} \lesssim D_{\nu_1} \lesssim 25 \, {\rm meV}$ for the lowest upper bound (black dashed line). 
Due to similar highest bound of the nEXO experiment, it also has possibility to explore the region of $20 \, {\rm meV} \lesssim D_{\nu_1} \lesssim 30 \, {\rm meV}$ (red dot-dashed line), while we may have a chance to constrain or find some signature in the region of $1 \, {\rm meV} \lesssim D_{\nu_1} \lesssim 15 \, {\rm meV}$ for its lowest upper bound (black dot-dashed line).

\subsection{IH}

In case of IH, we use the following BFs of the neutrino oscillation data:
\begin{align}
&\sin \theta_{12}^2 = 0.3088 \, , \quad \sin \theta_{23}^2 = 0.555 \, , \quad \sin \theta_{13}^2 = 0.02261 \, , \quad \delta_{\rm CP} = 283^{\circ} \, , \nn \\
&\Delta m_{\rm sol}^2 = 7.537 \times 10^{-5} \, {\rm eV}^2 \, , \quad \Delta m_{\rm atm}^2 \equiv D_{\nu_2}^2 - D_{\nu_3}^2 = 2.500 \times 10^{-3} \, {\rm eV}^2 \, .
\end{align}
%%%%%%%%%%%%%%%%%%%
\begin{figure}[tb]\begin{center}
\includegraphics[width=80mm]{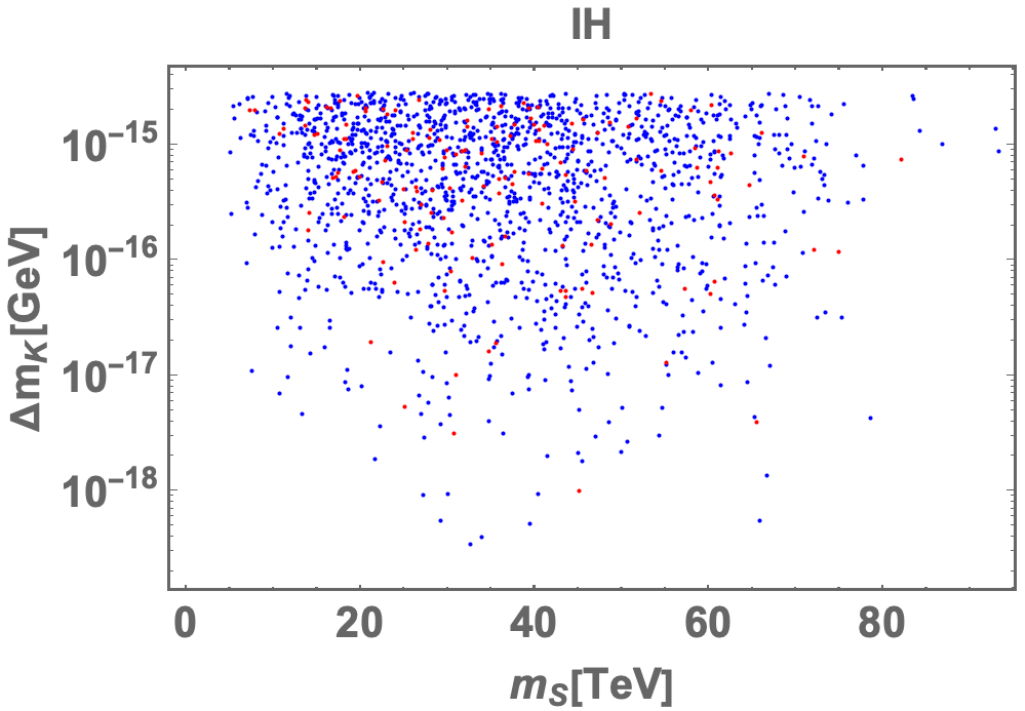}
\includegraphics[width=80mm]{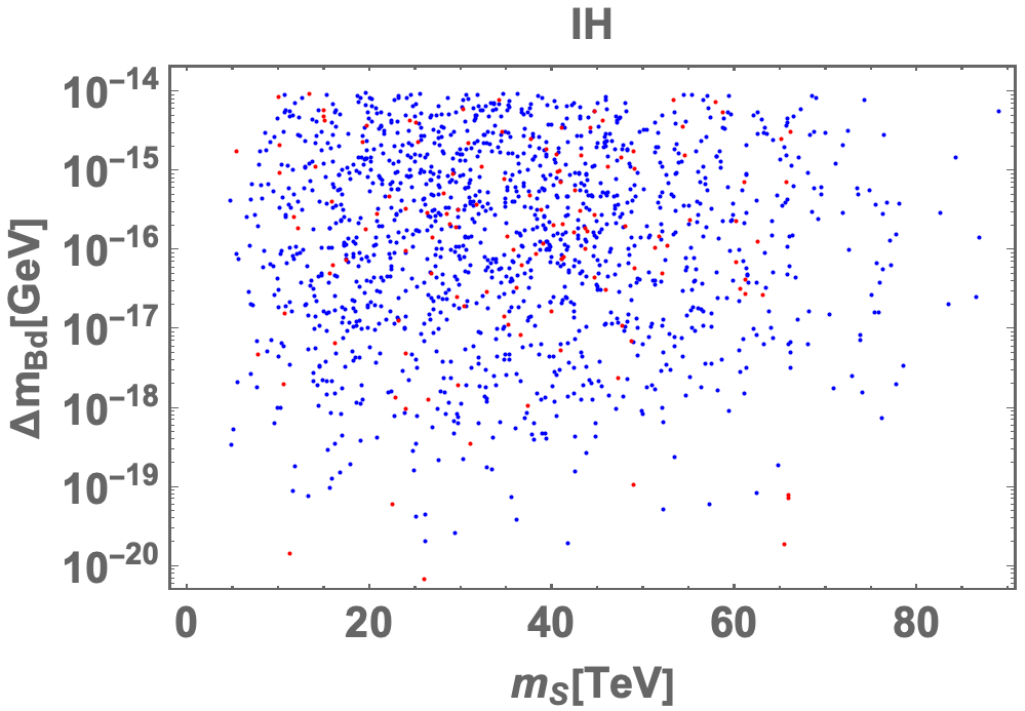}
\includegraphics[width=80mm]{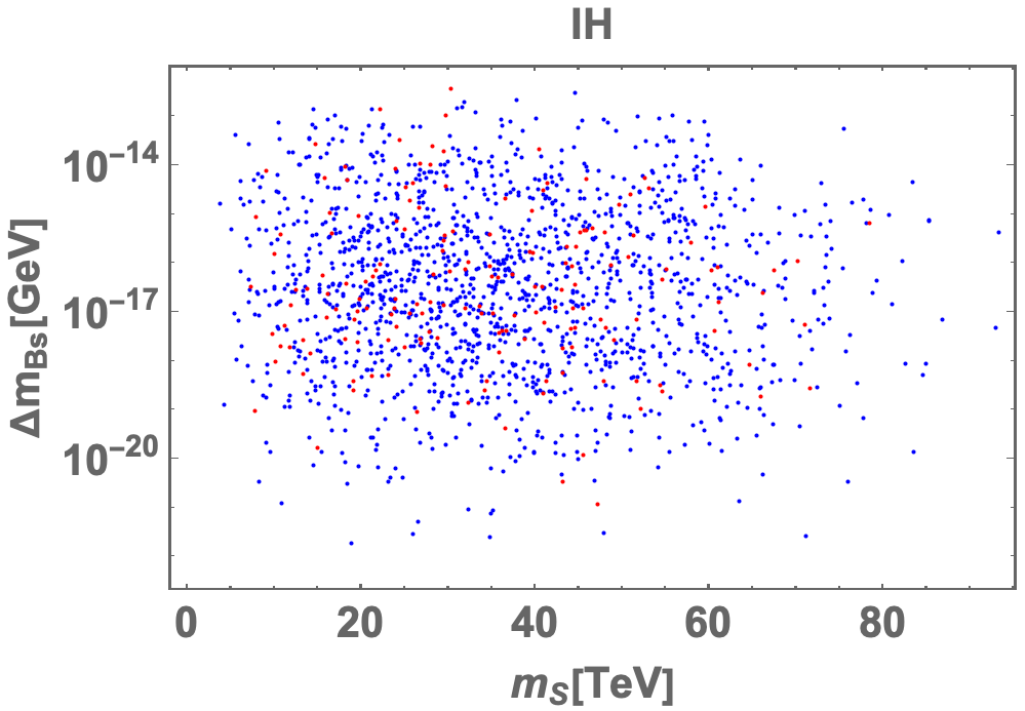}
\includegraphics[width=80mm]{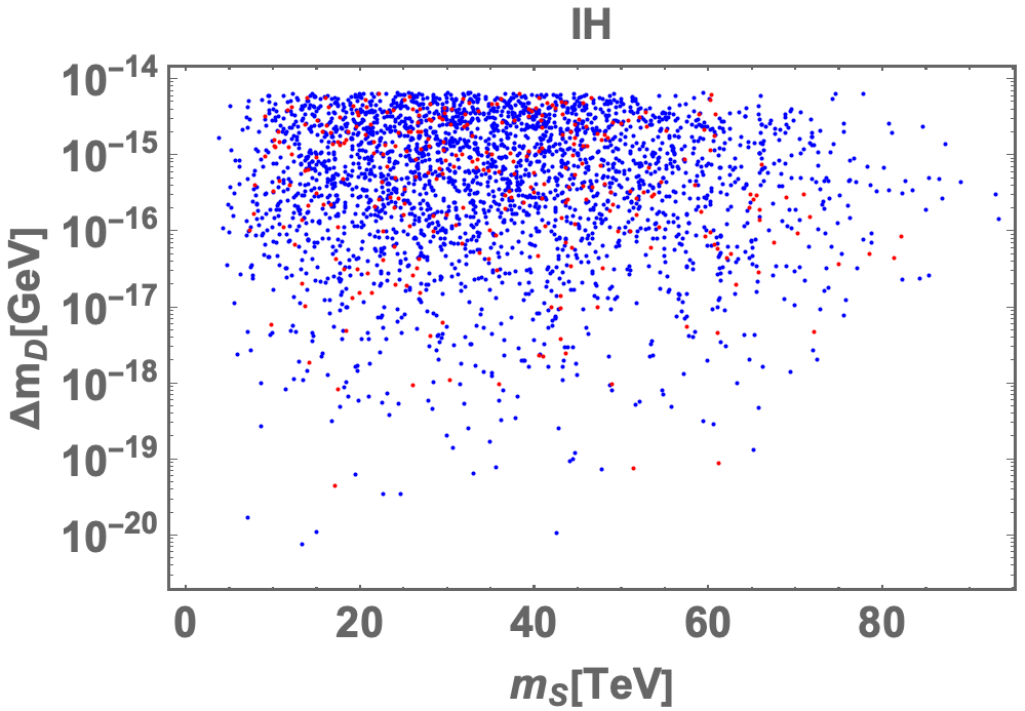}
\caption{Neutral meson mixings in case of IH, where all the legends are the same as the ones in Fig.~\ref{fig:nh1}. }
\label{fig:ih1}
\end{center}\end{figure}
%%%%%%%%%%%%%%%%%%%
In Fig.~\ref{fig:ih1}, we show our predictions of four neutral meson mixings with the same manner as in Fig.~\ref{fig:nh1}. 
These figures suggest that $\Delta m_{B_s}$ are the most weakest bound, and $\Delta m_{B_d}$ are the second weakest one. 
In contrast, $\Delta m_K$ ($\Delta m_D$) is the (second) most stringent bound, since a lot of allowed points are located at nearby each upper bound on our contributions. 
These tendencies are same as the case of NH in Fig.~\ref{fig:nh1}, and we can conclude that in our setup with one leptoquark $S$, it is not possible to distinguish the hierarchy of active neutrino masses from meson mixing parameters. 
Note that $m_S$ runs over wide range of $[4 \, {\rm TeV} \mathchar`- 95 \, {\rm TeV}]$, which is a bit wider than the NH case. 

%%%%%%%%%%%%%%%%%%%
\begin{figure}[tb]\begin{center}
\includegraphics[width=80mm]{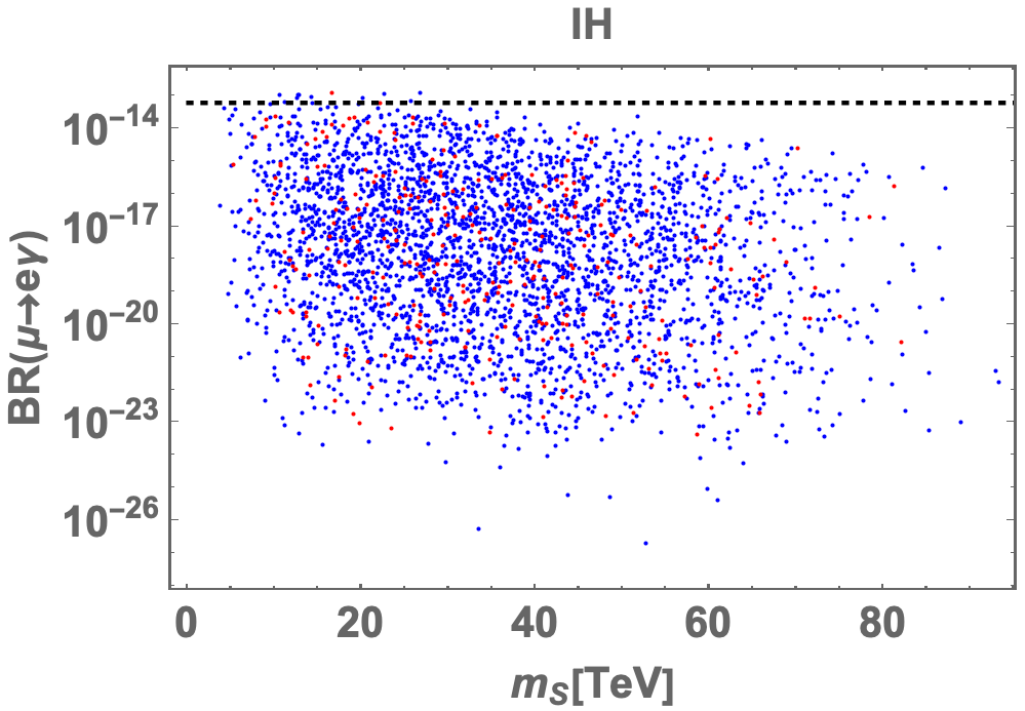}
\includegraphics[width=80mm]{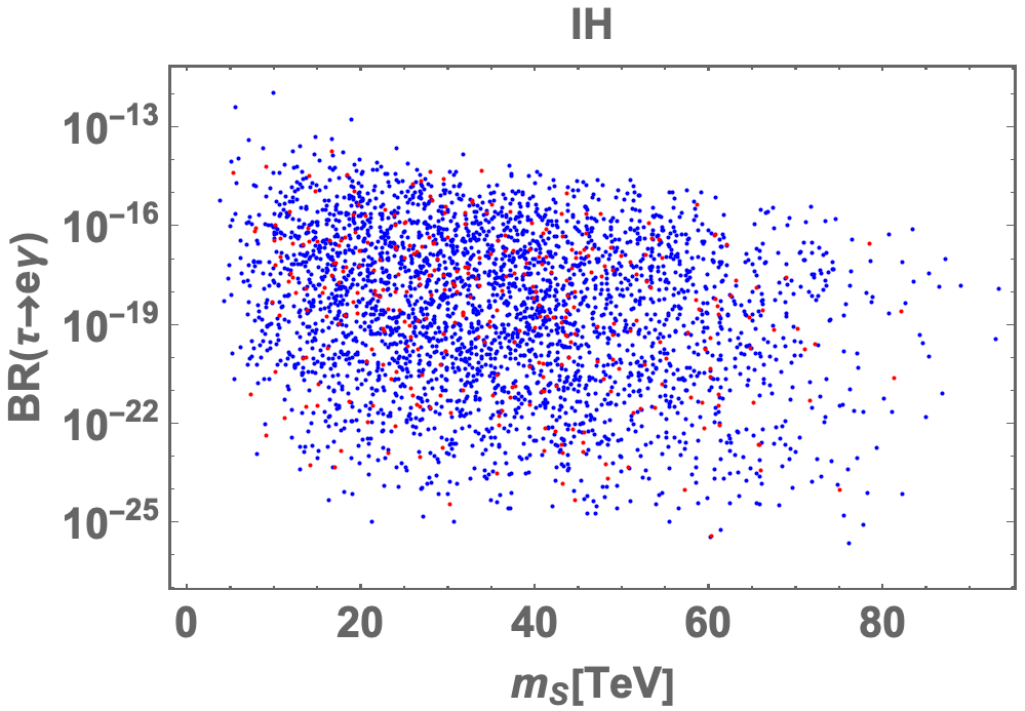}
\includegraphics[width=80mm]{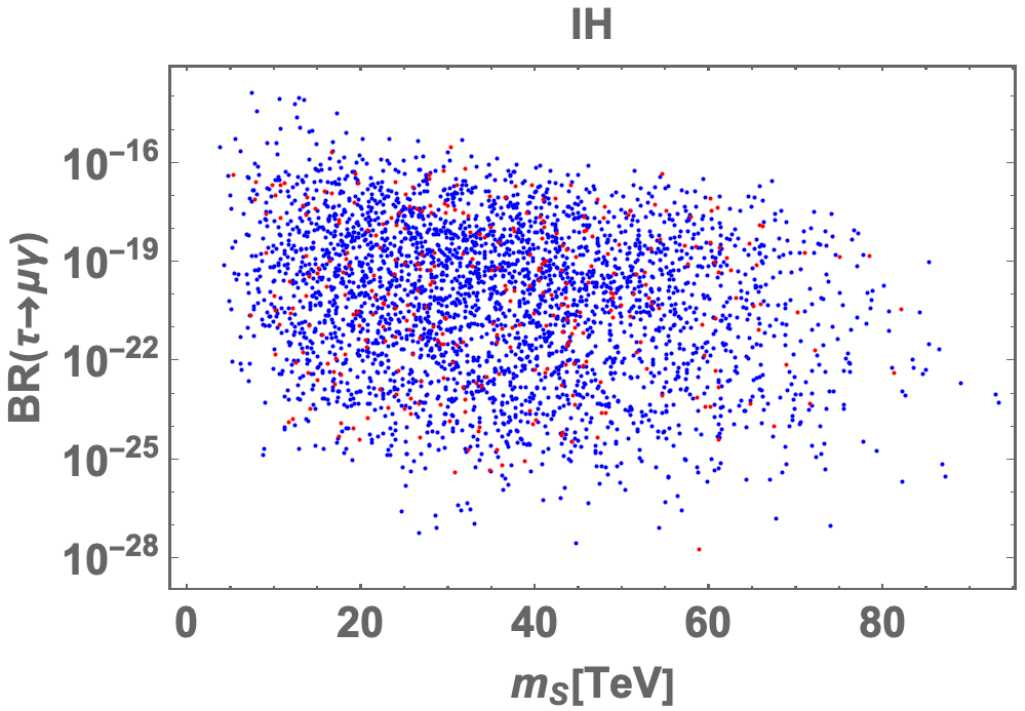}
\includegraphics[width=80mm]{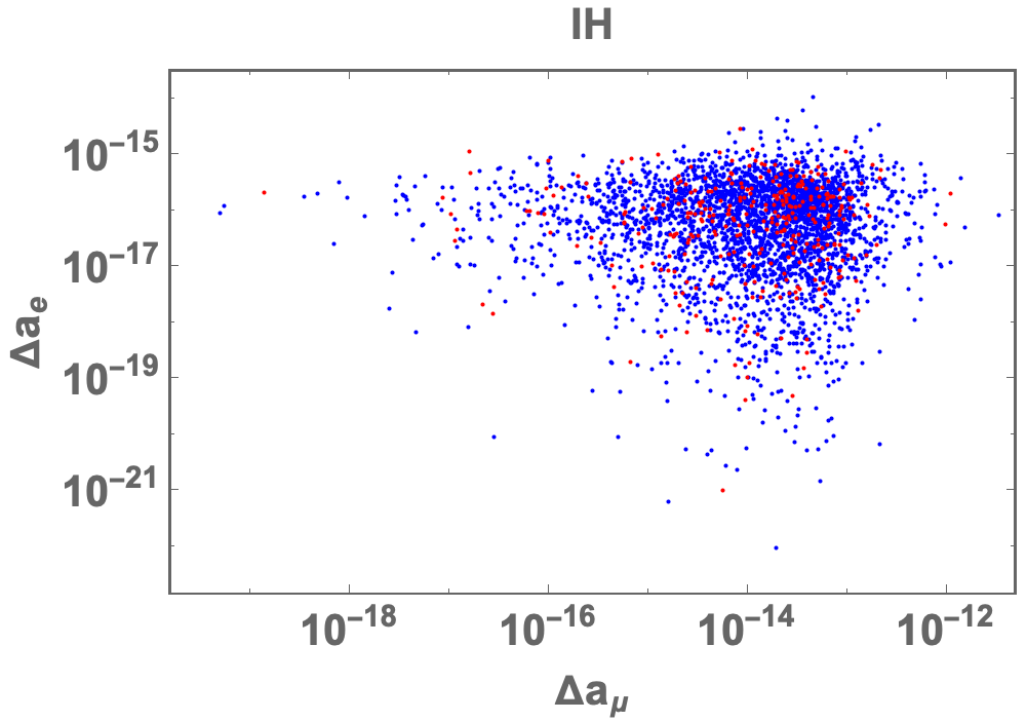}
\caption{LFVs and lepton $g-2$ in case of IH, where all the legends are the same as the ones in Fig.~\ref{fig:nh2}. }
\label{fig:ih2}
\end{center}\end{figure}
%%%%%%%%%%%%%%%%%%%
Next, in Fig.~\ref{fig:ih2}, we show our predictions of branching ratios for the three LFVs in terms of $m_S$, and $\Delta a_e$ versus $\Delta a_{\mu}$ plot, with the same ordering as in Fig.~\ref{fig:nh2}. 
The three branching ratios for the LFVs have similar tendencies to the case of NH, and the future experiment of $\mu \to e \gamma$ might exclude some points in our current region at $m_S = [7 \, {\rm TeV} \mathchar`- 27 \, {\rm TeV}]$, which is wider compared with the NH case. 
For the lepton $g-2$, the allowed points are located at specific region which is similar to the case of NH, although there are slightly smaller predictions for each $g-2$, e.g., $\Delta a_e \simeq 10^{-22}$ and $\Delta a_{\mu} \simeq 10^{-19}$. 

%%%%%%%%%%%%%%%%%%%
\begin{figure}[tb]\begin{center}
\includegraphics[width=80mm]{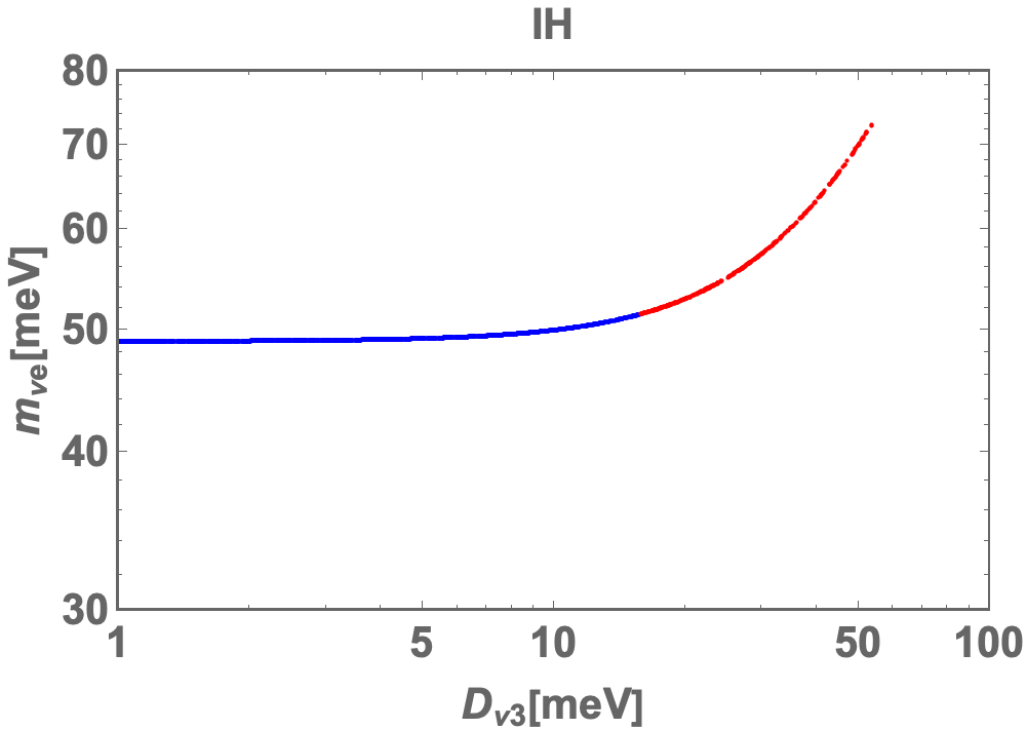}
\includegraphics[width=80mm]{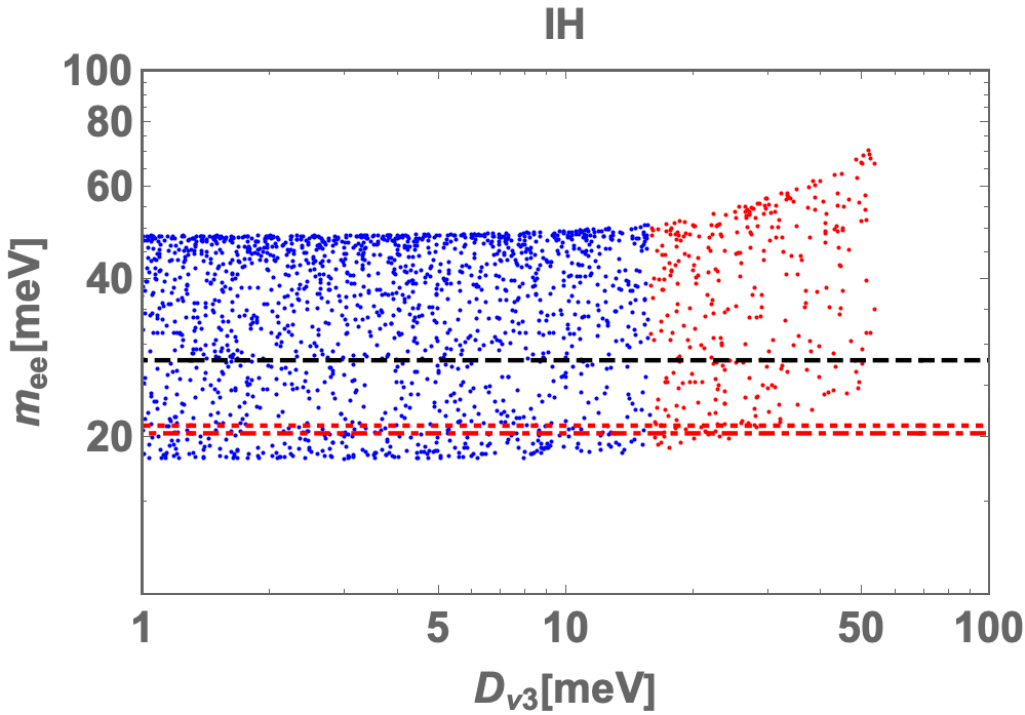}
\caption{Neutrino observations in case of IH, where all the legends are the same as the ones in Fig.~\ref{fig:nh3}. }
\label{fig:ih3}
\end{center}\end{figure}
%%%%%%%%%%%%%%%%%%%
In Fig.~\ref{fig:ih3}, we show $m_{\nu e}$ (left) and $m_{ee}$ (right) in terms of the lightest neutrino mass $D_{\nu_3}$. 
$m_{\nu_e}$ in the left panel results in the allowed range of $[49 \, {\rm meV} \mathchar`- 74 \, {\rm meV}]$, but it becomes narrower up to $51 \, {\rm meV}$ when $\sum D_{\nu} \le 120 \, {\rm meV}$ is taken into consideration. 
Therefore, we conclude that $m_{\nu_e} \simeq 50 \, {\rm meV}$ is favored for the IH case, which is larger than that for the NH case. 
We emphasize that nevertheless, the allowed region is still too small to reach the current experimental lower bound, $m_{\nu_e} \leq 450 \, {\rm meV}$. 
For $m_{ee}$ result in the right panel, we found that the allowed region is reached the current lowest upper bound, $28 \, {\rm meV}$ from the KamLAND-Zen experiment, which is shown by the black dashed horizontal line. 
However, it does not affect the lightest neutrino mass $D_{\nu_3}$ whose allowed region is up to $50 \, {\rm meV}$ ($17 \, {\rm meV}$) without (with) $\sum D_{\nu} \le 120 \, {\rm meV}$. 
Moreover, the highest upper bounds of the nEXO and the LEGEND-1000 experiments will severely exclude $m_{ee}$, and these bounds also restrict $D_{\nu_3}$ to be up to $\sim 30 \, {\rm meV}$ even when $\sum D_{\nu} \le 120 \, {\rm meV}$ is not imposed. 
It is clear that the lowest upper bounds of these experiments which are not shown explicitly in the plot cover whole allowed region, and hence, we may be able to find some signature or completely exclude our model with the case of IH.

%%%%%%%%%%%%%%%%%%%%
\section{Summary and discussion}
\label{sec:IV}

In this paper, we have proposed a minimal one-loop radiative  framework for the Dirac mass matrix, in which the active neutrino mass matrix is generated via the type-I seesaw mechanism, thereby alleviating the Yukawa hierarchies among the SM fermions. 
By introducing a scalar leptoquark and imposing appropriate assignments of ising fusion rule to the particle contents, we have successfully realized a minimal construction. 
Furthermore, the presence of the leptoquark leads to rich phenomenology, including semi-leptonic decays and neutral meson mixing as well as observable in the charged-lepton sector, i.e., lepton flavor violations and lepton $g-2$, rendering the model experimentally testable. 

After formulating each sector of our model, we have performed a comprehensive numerical analysis, taking into account all relevant experimental constraints for both normal and inverted hierarchies for active neutrino masses. 
Our analysis has revealed characteristic tendencies within the viable parameter space, as discussed in the numerical section. 
Moreover, we have discussed the testability of our allowed regions for both hierarchies and found that the majority of the parameter space can be probed by future experiments, such as nEXO and LEGEND-1000. 
In particular, for the case of IH, the entire allowed region lies within the future sensitivity limits of these experiments. 
Consequently, we expect either the detection of some signals from our model or a conclusion that our minimal setup favors the NH, based on the allowed values of $m_{ee}$. 

Before concluding, it is worthwhile to comment on the cut-off scale $\Lambda$. 
In our minimal realization, the introduction of an effective cut-off scale is necessary to regulate divergent loop contributions. 
One pragmatic approach is to associate this scale with the highest energy accessible at present or near-future experiments, namely $\sim 100 \, {\rm TeV}$. 
While this choice is phenomenologically motivated, it still involves a certain degree of theoretical ambiguity. 
A more fundamental determination of the cut-off scale from a theoretical perspective remains an important issue for future investigation.

%%%%%%%%%%%%%%%%%%%%%%%%%%%%%%%%%%%
\section*{Acknowledgments}

TN is supported by the Fundamental Research Funds for the Central Universities.
HO is supported by Zhongyuan Talent (Talent Recruitment Series) Foreign Experts Project.
YS is supported by Natural Science Foundation of China under grant No. W2433006. 
%%%%%%%%%%%%%%%%%%%%%%%%%%%%%%%%%%%

% Ref Style
% Including title
%\bibliographystyle{utphys}
\bibliography{ctma4.bib}
\end{document}